\documentclass{iucr}              

\usepackage{epsfig}
\usepackage{graphicx}
\usepackage{amsmath,amssymb}
\usepackage{color,xcolor}
\usepackage[normalem]{ulem}

\DeclareMathOperator{\sinc}{sinc}

     \journalcode{S}              

\begin{document}                  




\title{A hierarchical approach for modelling X-ray beamlines. Application to a coherent beamline.}


\cauthor[a]{Manuel}{Sanchez del Rio}{srio@esrf.eu}{}
\author[a]{Rafael}{Celestre} 
\author[a]{Mark}{Glass} 
\author[a]{Giovanni}{Pirro} 
\author[a]{Juan}{Reyes Herrera} 
\author[a]{Ray}{Barrett} 
\author[a]{Julio Cesar}{da Silva} 
\author[a]{Peter}{Cloetens} 
\author[b]{Xianbo}{Shi}
\author[b]{Luca}{Rebuffi}

\aff[a]{ESRF, 71 Avenue des Martyrs, 38043 Grenoble \country{France}}
\aff[b]{Argonne National Laboratory, 9700 South Cass Avenue, Lemont, Illinois 60439, \country{USA}}

%
%
%








\maketitle                        

\begin{synopsis}
A hierarchical scheme of the calculations of an X-ray beamline for coherent applications is presented. Starting from simple calculations to partial coherence calculations, we show how ray-tracing and wave optics software are applied simulating the beamline performances.   
\end{synopsis}

\begin{abstract}
We consider different approaches to simulate a modern X-ray beamline. Several methodologies with increasing complexity are applied to discuss the relevant parameters that quantify the beamline performance. Parameters such as flux, dimensions and intensity distribution of the focused beam and coherence properties are obtained from simple analytical calculations to sophisticated computer simulations using ray-tracing and wave optics techniques. A latest-generation X-ray nanofocusing beamline for coherent applications (ID16A at the ESRF) has been chosen to study in detail the issues related to highly demagnifying synchrotron sources and exploiting the beam coherence. The performance of the beamline is studied for two storage rings: the old ESRF-1 (emittance 4000~pm) and the new ESRF-EBS (emittance 150~pm). In addition to traditional results in terms of flux and beam sizes, an innovative study on the partial coherence properties based on the propagation of coherent modes is presented. The different algorithms and methodologies are implemented in the software suite OASYS \cite{codeOASYS}. Those are discussed with emphasis placed upon the their benefits and limitations of each.
\end{abstract}

\section{Introduction}

Many storage-ring-based X-ray synchrotron facilities are building or planning upgrades to increase brilliance and coherent flux by one to three orders of magnitude.  The first upgrade of a large facility will be the EBS (Extremely Brilliant Source) \cite{orangebook} at the European Synchrotron Radiation Facility (ESRF). The new storage ring of 150~pm emittance will be commissioned in 2020 and will significantly boost the coherence of the X-ray beams. Applications exploiting beam coherence such as X-ray photon correlation spectroscopy, coherent diffraction imaging, propagation-based phase-contrast imaging, and ptychography will strongly benefit from this update.  

Accurate calculation and quantitative evaluation of the parameters related to X-ray optics are of great importance for designing, building and exploiting new beamlines. Every modern beamline follows a procedure of conceptual design plus technical design where detailed simulations are essential. These calculations allow testing of the design parameters in a virtual computer environment where advantages and limitations can be studied quantitatively. The selection of the optimal optical layout requires a detailed simulation of the optics imperfections, that in many cases are the limiting factor of the beamline optics. Deformations of the optical elements due to heat load need to be controlled, therefore optics simulations have to include results from engineering modeling of the thermal deformations usually done using finite element analysis. Traditionally the optics calculations for synchrotron beamlines are done using ray-tracing techniques. They model the X-ray beam as a collection of incoherent rays. This technique is still essential for obtaining information on aberrations, flux propagation, monochromator bandwidth, etc. However, the high coherence of the new sources can introduce the necessity to complement the ray-tracing methods with models including effects of diffraction and scattering from a coherent beam. A fully coherent hard X-ray beam is still a dream for the new generation storage rings. Despite a considerable increase in the coherent fraction of the new beams, the coherent fraction at typically hard X-ray photon energies is still a few percent for the new sources. Consequently, the ray-tracing analysis does not become obsolete but has to be complemented with an analysis based on wave optics. Because of the partial coherence, the wave optics analysis becomes nontrivial and can be done with different levels of approximation, as discussed in this work. In addition, wave optics simulations are usually computationally expensive.

To illustrate the applicability of a variety of methods and solutions for beamline optical simulations we have chosen the ESRF beamline ID16A~\cite{ID16A}. This beamline provides a high-brilliance beam focused to a few nanometers. It displays characteristics increasingly aimed for in modern beamlines: extremely high demagnification (sub-micrometer focused beam size) and high coherence. It combines coherent imaging techniques and X-ray fluorescence microscopy to perform quantitative 3D characterization of the morphology and the elemental composition of specimens in their native state~\cite{XRFID16A}.

X-ray beamlines are very particular optical systems. Some implications of exploiting the short wavelengths such as the use of grazing incidence reflectors and the use of perfect crystals as typical hard X-ray monochromators render common commercial software packages ill-adapted for the study of synchrotron beamlines. For this reason, many optics tools have been developed by the synchrotron community. Some of them have a long history like SHADOW - from 1984 \cite{firstShadow}; or SRW - from 1998 \cite{codeSRW}; to mention only codes used in this work. These historical tools together with other new applications have been collected together in a user-friendly software package: OASYS \cite{codeOASYS}. Many of the calculations presented here have been performed directly in OASYS or with scripts created with the help of OASYS. 

 In this paper, some quantitative results on how the storage ring upgrade will affect a particular beamline are presented. The paper is structured as follows: we first introduce in Section~\ref{Beamline description} the main parameters of the beamline and explain the particular choice of the parameters used in the calculations. Section~\ref{Simulations} is the central part of the paper, and describes the different methods used to analyze the beamline with increasing complexity, setting a hierarchy of methodologies that can be exported to the simulations of any beamline. In Section \ref{level0} we calculate with simple methods approximate values of focal size, flux and coherent fraction. Accuracy can be improved using geometrical methods in Section \ref{level1} that allow a more complete simulation of the beamline optics within certain limits which are related, in this example, to high demagnification and high coherence of the studied beamline. The wave optics section~\ref{level2} includes different models of increasing complexity as the required accuracy of the results becomes more demanding. In the Discussion (Section \ref{discussion}) the results of the different simulation methods are commented. The hierarchical approach allows more accurate and better results but the price to pay is to deal with more complex and computationally expensive tools. Finally, Section \ref{summary} summarizes the results.

 %

\section{Beamline description}
\label{Beamline description}

This section presents the reader with the main operational parameters for the ID16A beamline at ESRF \cite{ID16A} in two scenarios. This beamline was built during the Upgrade Phase~I at ESRF (2009-2015). It uses one or several undulators as the source. During the period 2011-2018, it operated exploiting the ESRF-1 storage ring’s double-bend achromat (DBA) lattice with horizontal emittance about 4000~pm. In the second phase of the ESRF Upgrade, the original storage ring is being replaced by a new hybrid multi-bend achromat (HMBA) lattice. This new lattice will drastically improve the performances of the ESRF source. The resulting Extreme Brilliant Source (EBS) \cite{orangebook} will increase the brightness and coherent fraction by a factor of more than 100, by strongly reducing the horizontal emittance. This is possible using almost the same ring circumference and electron energy $E_e$. As a result, the X-ray sources will be much more coherent, especially in the horizontal direction.

ID16A is a beamline that provides a nanometric, highly coherent beam with broad energy-bandpass. It is dedicated to nanofocusing applications. The beamline can operate at two fixed photon energies: 17050~eV and 33600~eV. The straight section in the magnet lattice houses four undulators, two 18~mm and another two of 22.4~mm period. The undulator parameters and $K$-values at the resonances for the energies in use are shown in Table \ref{tab:Undulators}. 

\begin{table}\label{tab:Undulators}
\centering
\caption{Parameters of the undulators available at the ID16A beamline: period ($\lambda_u$), length ($L$), number of identical undulators ($N$) and deflection parameter ($K$) values for the three main photon energies in use. The valuse are shown for the two ESRF magnetic lattices: the old ESRF-1, with $E_e = 6.04$~GeV), and the new ESRF-EBS, with $E_e = 6.00$~GeV. $n$ is the emission harmonic in use for each particular photon energy.}

\resizebox{\textwidth}{!}{%
\begin{tabular}{lcccccc}
name  & $\lambda_u$ [mm] & $L$ [m] & N & $K$ (11200~eV) & $K$ (17050~eV) & $K$ (33600~eV) \\ \\
 ESRF-1 U18.3  & 18.30  & 1.40  & 2 & 1.175(n=1) & 0.470(n=1) & 1.175(n=3)\\ 
 ESRF-1 U22.4  & 22.40  & 1.40  & 2 & 0.873(n=1) & 1.855(n=3) & 0.873(n=3)\\ \\
 EBS U18.3  & 18.30  & 1.40  & 2 & 1.156(n=1) & 0.437(n=1) & 1.156(n=3)\\ 
 EBS U22.4  & 22.40  & 1.40  & 2 & 0.852(n=1) & 1.836(n=3) & 0.852(n=3)\\ 
\end{tabular}%
}
\end{table}

\begin{figure}\label{fig:ID16A}
    \centering
    \includegraphics[width=0.9\textwidth,clip=true,trim=50 85 50 85]{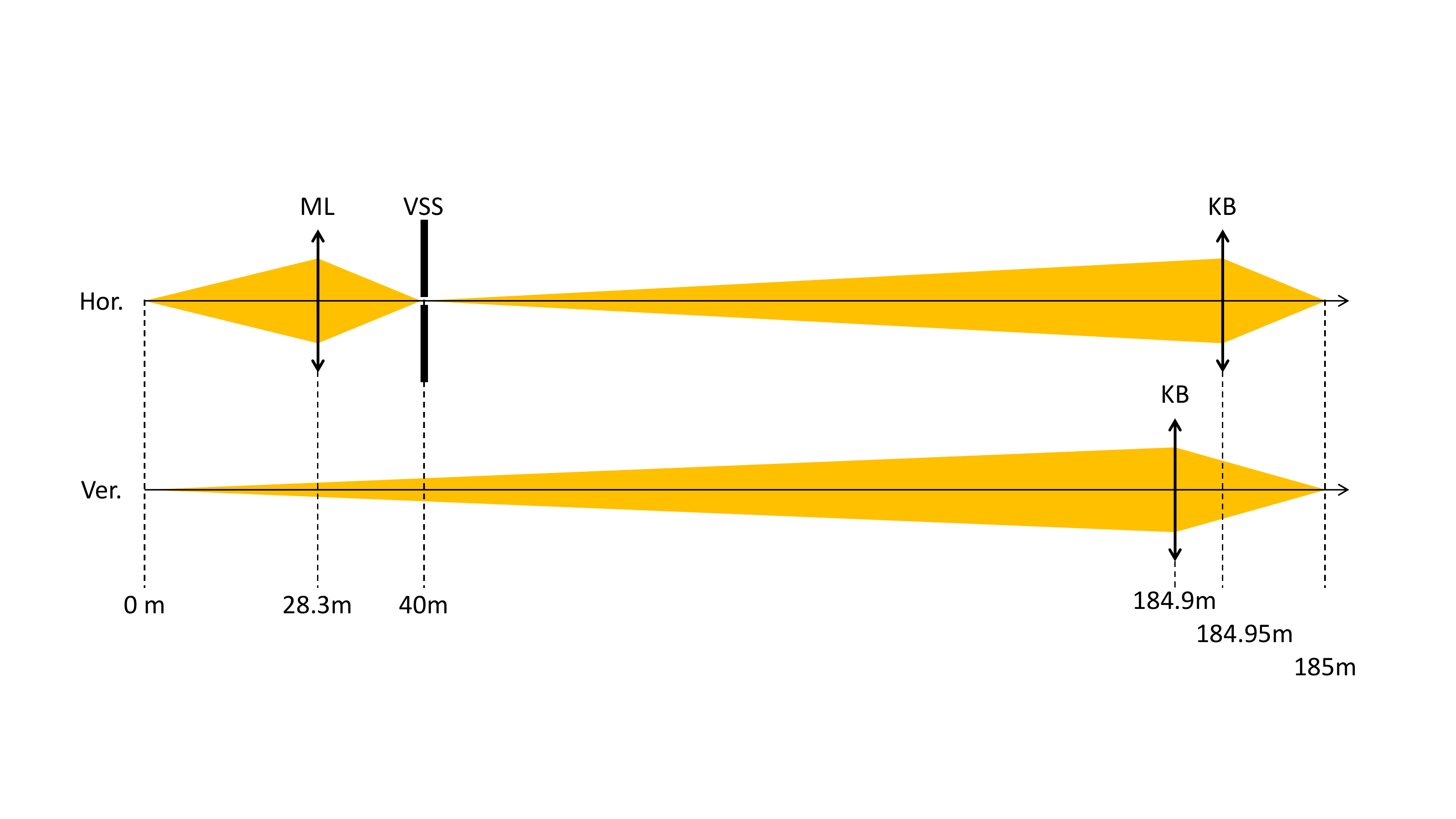}
    \caption{Schematic of the ID16A beamline showing the horizontal (top) and vertical (bottom) planes and the position of the main optical elements. In the sketch ML stands for the focusing multilayer monochromator; VSS for virtual source slit; and KB for the Kirkpatrick-Baez set of mirrors. The sketch is drawn out of scale.}
\end{figure}

ID16A is a long beamline: the distance from the center of the straight section to the sample position is 185~m. Strong focusing is performed using a Kirkpatrick-Baez (KB) mirror system placed a few cm upstream of the sample position (focal plane), which ensures a high demagnification. In order to match the acceptance of the KB in the vertical and horizontal planes, a focusing (cylindrically bent) multilayer monochromator (ML) has
been chosen. It i) monochromatizes the beam with a large energy bandwidth (typically $\Delta E/E \approx 10^{-2}$) to allow high flux, 
and ii) it focus tangentially the beam in the horizontal plane to create a secondary source at 40~m downstream of the undulator, where a horizontal slit (VSS) is placed. A beamline schematic is shown in Figure~\ref{fig:ID16A}. The positions of the different elements are shown in Table \ref{tab:Positions}. This table also shows the extremely high geometrical demagnification ($\sim 10^3$) needed for focusing the beam to the nanometric range.  

\begin{table}\label{tab:Positions}
    \caption{Position of the main ID16A optical elements with respect to the source (undulator centre). It also shows the values for horizontal and vertical demagnification $M^{-1}=p/q$, where $p$ and $q$ are the the distances object-optics and optics-image, respectively.}
    \resizebox{\textwidth}{!}{%
    \begin{tabular}{rcccccc}
    -                      & Source & Multilayer (ML) & Slit (VSS) & KB (v) & KB (h) & Focal plane \\ 
    Position {[}m{]}:      & 0.00   & 28.30      & 40.00 & 184.90 & 184.95 & 185.00      \\
    Horizontal $M^{-1}$:   & -      & 2.42       & -     & -      & 2899   & -           \\
    Vertical $M^{-1}$:     & -      & -          & -     & 1849   & -      & -           \\ 
    \end{tabular}%
    }%
\end{table}


\begin{table}\label{tab:TableSources}
    \caption{Main parameters of the present DBA lattice (ESRF-1) and future HMBA lattice (EBS) storage rings \cite{orangebook}.}
        \begin{tabular}{rcc}
                                              & \textbf{ESRF-1} & \textbf{EBS}      \\ 
        Lattice type:                         & DBA           & HMBA              \\
        Circunference {[}m{]}:                & 844.930       & 843.979           \\
        Beam energy {[}GeV{]}:                & 6.04          & 6.00              \\
        Beam current {[}mA{]}:                & 200           & 200               \\
        Natural emittance {[}pm$\cdot$rad{]}: & 4000          & 147               \\
        Energy spread {[}$\%${]}:             & 0.0011        & 0.00093           \\       
    \end{tabular}
\end{table}

\begin{table}\label{tab:eBeam}
    \centering
    \caption{ESRF-1 and EBS main electron beam parameters. Electron beam parameters for the high-$\beta$ (labelled ESRF) and the EBS straight sections. Values taken at the symmetry point of the straight section, where the insertion devices are placed. Values are taken from \cite{ESRF2014}, updated for EBS with
    the lattice labelled S28D. 
    }
        \resizebox{\textwidth}{!}{%
        \begin{tabular}{ccccc}
         Lattice & $\sigma_e$ horizontal {[}$\mu$m{]} &  $\sigma_e'$ horizontal {[}$\mu$rad{]} & $\sigma_e$ vertical {[}$\mu$m{]} & $\sigma_e'$ vertical {[}$\mu$rad{]} \\ \\
         ESRF-1  & 387.80  & 3.50  & 10.30 & 1.20 \\ 
         EBS     & 30.18  & 3.64  & 4.37 & 1.37 \\ 
    \end{tabular}
    }
\end{table}

\section{Simulations}
\label{Simulations}

For all simulations presented hereafter, we selected the configuration using a single undulator U18.3 placed in the center of the straight section and tuned to its first harmonic at a photon energy of $E=17225$~eV: a value close to the usual operation value shown in Table \ref{tab:Undulators}. Calculations were done using the two different storage ring lattices mentioned before: the (already dismantled) ESRF-1 where the undulators are in a high-$\beta$ straight section (larger horizontal emittance) and the new EBS (lower horizontal emittance) straight sections. The parameters for the current and the new lattice are shown in Tables \ref{tab:TableSources} and \ref{tab:eBeam}.
This section is divided into three subsections, that present three main levels of simulations: simple calculations that can be made by hand, ray-tracing simulations and wave optics simulations: sections  \ref{level0}, \ref{level1} section \ref{level2} respectively.   

\subsection{Analytical calculations}
\label{level0}

We first estimate the beam size at different positions in the beamline, in particular at the source plane and at the image plane. A first na{\"{i}}ve estimation of the focal size can be made by considering the source size times the magnification factor $M$. In the paraxial approximation, the magnification factor can be expressed as $M=q/p$, where $p$ is the distance between object and imaging element and $q$ is the distance between the imaging element and image. The demagnification is the inverse of the magnification ratio: $M^{-1}=p/q$. Systems where $M>1$ are said to be magnifying, whereas systems with $M<1$ (or $M^{-1}>1$) are said to be demagnifying. The latter will be our case: it is, consequently, more convenient to work with demagnification ($M^{-1}>1$) rather than with magnification. 

The source size and divergence are usually approximated supposing a Gaussian distribution of the electron positions and divergences (Table \ref{tab:eBeam}) and single-electron photon source size and divergence. In some cases, such as for undulator emission, this Gaussian approximation may not be accurate and have some implications that will be discussed later. Let $\sigma_{e(h,v)}$ and $\sigma_{e(h,v)}'$ be the standard deviations of electron beam size and divergence at the center of symmetry of the straight section in a plane perpendicular to the main propagation direction. The subscripts $h$ and $v$ stand for horizontal and vertical, respectively. 

The undulator natural photon source size and divergence are given by \cite{elleaume}:
\begin{align}
    \label{eq:photon small sigmas}
    \sigma_u=\frac{2.74}{4\pi}\sqrt{\lambda L}\approx \frac{1}{2  \pi}\sqrt{2 \lambda L}  && \sigma_u' = 0.69\sqrt{\frac{\lambda}{L}}\approx \sqrt{\frac{\lambda}{2 L}},
\end{align}
where $\lambda$ is the emitted wavelength and $L=N\Lambda_u$ is the undulator length, with $\Lambda_u$ as the magnetic period of the undulator and $N$, the number of magnetic periods. The electron beam usually presents different horizontal and vertical beam emittances, but the undulator natural source size or divergence presents radial symmetry at the central cone (excluding polarization). The photon source size and divergence are approximately given by the convolution of the electron beam sizes and the undulator natural sizes. If we consider Gaussian distributions, the photon source has sizes $\Sigma$ and divergences $\Sigma'$ given by:
\begin{align}
\label{eq:photon big sigmas}
\Sigma_{h,v}=\sqrt{\sigma_{e(h,v)}^2 + \sigma_u^2} && \Sigma_{h,v}'=\sqrt{\sigma_{e(h,v)}'^2 + \sigma_u'^2}.
\end{align}

The beam sizes at different positions in the beamline are estimated in different ways depending upon whether the beam is focused or not. If the beam is focused, its dimensions, to a first approximation, are the photon source size multiplied by the corresponding magnification factor. For predicting the photon beam sizes in a position out of the focus the beam divergence is used. In an ideal {\'{e}}tendue-conserving optical system, where the Smith-Helmholz invariant applies, the resulting beam divergence transforms inversely as the beam size: it can be calculated as the product between the beam divergence at the source with the demagnification factor $M^{-1}$.

Using the values for U18.3, $n$=1, and  $E$=17225~eV (Table \ref{tab:Undulators}) and the optical configuration described in Table \ref{tab:Positions}, one obtains a focal size 139$\times$5~nm$^2$ (H$\times$V) for ESRF-1 and $10\times5$~nm$^2$ for EBS (Table~\ref{tab:HandCalculations}). The horizontal focal spot for the ESRF-1 lattice is larger than the specified target value for the beamline, therefore the VSS could be closed to 50~$\mu$m thus reducing the beam size at the VSS plane by a factor ca. 8.
Therefore, for ESRF-1, the horizontal size at the sample will be affected by the same factor obtaining roughly 17$\times$5~nm$^2$.

In general, for simple elliptical cylinder figured grazing incident optics imaging extended sources with large demagnifications ($M^{-1} \gg 100$), the effect of optical aberrations is very important even using a perfect optical surface. As a consequence, the results calculated using only the demagnification factor are usually optimistic. Moreover, imperfections in the optical surface (figure errors, slope errors and microroughness) are often the limiting factors of real reflective optics. Their effects will be studied in Sections~\ref{level1} and \ref{level2}.


\begin{table}\label{tab:HandCalculations}
    \centering
    \caption{Photon beam size and divergence for selected ID16A beamline positions. Values below were obtained considering undulator U18.3 tuned to its first harmonic at the photon energy of $E=17225$~eV for both the high-$\beta$ (ESRF-1) and the EBS straight sections. Values are FWHM. Values in parentheses correspond to closing the VSS slit to 50~$\mu$m.
    }
    \resizebox{\textwidth}{!}{%
    \begin{tabular}{r|ccc|ccc}
                            &               & \textbf{ESRF-1 (high-$\beta$)} &   &         & \textbf{EBS}  &    \\ 
    Element:                & Undulator     & VSS          & Sample            & Undulator & VVS   & Sample       \\
    Position [m]:           & 0.00          & 40.00        & 185.00            & 0.00    & 40.00   & 185.00       \\
    FWHM$_h$ [$\mu$m]:      & 977.19        & 404.00 (50.0)& 0.139 (0.017)     & 71.26   & 29.46   & 0.0102       \\
    FWHM$_v$ [$\mu$m]:      & 9.60          & 482          & 0.0052            & 10.02   & 486.5   & 0.0054       \\
    \newline
    FWHM'$_h$ [$\mu$rad]:   & 26.96         & 65.22        & 189062   & 15.60   & 37.74   & 109404       \\
    FWHM'$_v$ [$\mu$rad]:   & 12.04         & 12.04                 & 22264    & 12.16   & 12.16   & 22489        \\                                                   
    \end{tabular}%
    }
\end{table}

\begin{figure}\label{fig:FluxU18}
    \centering
    \includegraphics[width=0.95\textwidth]{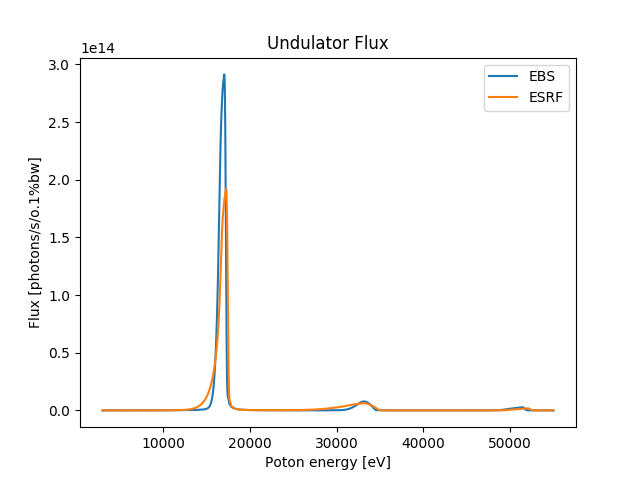}
    \caption{Flux of the U18.3 undulator integrated over a square slit placed at 27.1~m from the ID centre with aperture of 1 and 4 mm.}
\end{figure}

Next, we want to estimate the number of photons in the sample. For that, we need the spectrum at the source. It is not straightforward to get this information analytically but one can calculate the flux and power using different codes. Perhaps the most advanced and well-maintained codes used in the synchrotron community are SPECTRA \cite{codeSPECTRA} and SRW \cite{codeSRW}. The XOPPY package in the OASYS suite implements three codes: US \cite{codeUS}, URGENT \cite{codeURGENT} and SRW. Figure~\ref{fig:FluxU18} shows the spectra calculated using SRW over an aperture of $1 \times 1$~mm$^2$ placed at $d=27.1$~m. This aperture opening is set to fully accept the central cone at the resonance peak of the harmonic in use.
The peak intensity is about  $2.9\cdot10^{14}$ photons/s/0.1{\%}bw for EBS and $1.9\cdot10^{14}$ for ESRF-1 (see \ref{fig:FluxU18}). One can appreciate how the emittance plays an important role in both peak width and especially in peak value. The number of photons per second after the multilayer monochromator is $2.9\cdot10^{15}$ for EBS and $1.9\cdot10^{15}$ for ESRF-1, considering that the energy bandwidth transmitted by a multilayer monochromator is $\Delta E/E \approx 10^{-2}$ and the bandwidth in Fig.~\ref{fig:FluxU18} is 10$^{-3}$, and supposing, at this point, ideal reflectivity (100~\%) for the multilayers. 

To estimate the geometrical transmission of the beamline we have to consider which optical elements crop the beam and remove photons. This can be analyzed using their angular acceptance or numerical aperture. The multilayer monochromator accepts the whole beam. We consider a VSS closed to $S_h$=50~$\mu$m in horizontal and fully opened in vertical. For ESRF-1 its transmission is 1/8 (as discussed before) and 1 for EBS. In a similar way, the KB mirrors will crop the beam to a size that is the projection of their useful optical length onto a direction perpendicular to the optical axis. Using a mirror length of $L_v$=60~mm and $L_h$=26~mm for VFM and HFM, respectively, and incident angle of 15~mrad, the KB behaves as an aperture of 900 $\times$ 390~$\mu$m$^2$. We get N.A. values of 6.2 $\times$ 21.6~$\mu$rad$^2$ (aperture size divided by the distance from the mirror to the VSS in H and to the source in V). The transmission coefficient is the ratio of the accepted N.A. over the beam divergence, resulting in 9.5\% (H) and 0.18\% (V)for ESRF-1 and 16.3\% and 0.18\% for EBS. The overall transmission is the product of H and V ratios, therefore 0.2\% for ESRF-1 and 2.95\% for EBS, therefore $3.8\cdot10^{12}$ photons/s for ESRF-1 and $8.5\cdot10^{13}$ photons/s for EBS. 

One can also make some simple inferences about the beam coherence. The coherent fraction (CF) \cite{arxivCF} of the undulator beam is the occupation of the lowest coherent mode and will be discussed further in Section~\ref{level2}. An approximated estimation of CF can be obtained using Eqs.~\ref{eq:photon small sigmas} and \ref{eq:photon big sigmas}. The coherent fraction is the ratio of the phase space of the fully coherent (zero emittance) beam relative to the true beam. The CF in the horizontal and vertical planes are
\begin{equation}\label{eq:coherent fraction}
 CF_{(h,v)} = \frac{\sigma_u \sigma'_u}{\Sigma_{(h,v)} \Sigma'_{(h,v)}},
\end{equation}
and the total coherent fraction is the product of both (CF=CF$_h \times $CF$_v$). An estimation of the CF calculated in this way gives CF=0.12\% for the ESRF-1 and 2.73\% for EBS at 17 keV photon energy. The CF roughly indicates the ratio of the ``coherent photons'' over the total number of photons. It is a good indicator of the quality of a coherent beam (a fully coherent beam has CF=1 and an incoherent beam CF=0). The beamline optics act as a ``coherence filter'' by removing the ``bad'' or ``incoherent'' photons and in consequence increasing the CF. A complete cleaning is impossible because of the spatial overlapping of the different coherence modes. However, apertures centered at the optical axis would clean the beam from these undesired photons. If we compare the values of the beamline transmission due to the geometrical aperture  (0.2\% for ESRF-1 and 2.95\% for EBS) with the CF values, we remark that they are very close. This suggests that the beam at the focal position will be highly coherent because the fraction of the photons removed from the beam matches the CF value that quantifies the fraction of ``good'' or ``coherent'' photons. 

To assess the focused beam size given by a coherent beam, one can consider the beamline in two sections: First, the ML monochromator and VSS that produce a coherent plane wave as the source (vertical) and VSS (horizontal) are both very far from the entrance aperture of the focusing optics. Second, the KB-system playing a dual role, acting as a focusing element of focal distance $f=5$~cm, but also cropping the beam by acting as a square aperture of roughly D$\times$D=400$\times$400~$\mu$m$^2$ (the length of the mirror projected in a plane perpendicular to the beam propagation).
This aperture has two effects: i) as discussed before, it removes a lot of off-axis photons increasing the CF to a value close to one, and ii) broadens of the focal spot because of coherent diffraction. The diffraction by a square aperture of a collapsing spherical wavefront has an intensity distribution proportional to $\sinc^2(k D x/2 f) \sinc^2(k D y / 2 f)$,
where $k=2\pi/\lambda$, and $x$ ($y$) is the horizontal (vertical) coordinate. Considering that the FWHM of $\sinc(x)$ is approximately 2.78, one obtains a FWHM of 7.75$\times$7.75~nm$^2$ for the focal spot. This spot size is of the order or larger than that calculated geometrical considerations alone and therefore contributes significantly to the overall size of the focused beam. One can conclude that for the conditions in use this is a diffraction-limited beamline.

In summary, we have shown in this section how simple analytical calculations that can be done mostly ``by hand'' help to estimate the geometrical beam size at focal position (17$\times$5~nm$^2$ for ESRF-1, 10$\times$5~nm$^2$ for the EBS) the flux ($3.8\cdot10^{12}$ photons/s for ESRF-1 and $8.5\cdot10^{13}$ photons/s for EBS) and also anticipate that the beam is highly coherent producing a diffraction-limited focal spot of  7.75$\times$7.75~nm$^2$. Thus far we ignored the effect of real optics exhibiting surface errors, the reflectivity of the elements, and other aspects that will be treated in depth in the next sections. This is, however, the first step to be done when analyzing a beamline. It also helps in providing an initial benchmark estimate which should, to a large extent, provide a check of the validity of the numerical calculations that will provide more precise values, but should remain at the same order of magnitude.  

%
\subsection{Ray-tracing calculations}
\label{level1}
In this section, we perform ray-tracing calculations to study the effect of real optics (including aberrations and slope errors) upon the beam size and beamline transmission. The first part deals with the effect of aberrations and transmission due to geometrical considerations. The second part shows how the effect of slope errors of the KB mirrors degrade the focal spot.  

Ray-tracing is a simulation method based on tracing some light rays along the optical system from the source to the image and retrieving the statistics of rays at the image to measure beam characteristics. For simple systems one can perform ray-tracing even by hand, selecting some principal rays that will define the location and envelope of the image. Using computers one can trace thousands of rays and calculate the position and divergence distribution of the rays using statistics. Rays are usually generated by Monte Carlo sampling the source characteristics. Ray-tracing is based on geometrical optics: a ray is a solution of the Helmholtz equation and travels in vacuum along a straight line. Rays are intercepted by the optical elements, which change their trajectory. For reflectors, this change of direction is given by the specular reflection laws. For refractors, the change of direction is defined by  Snell's law. Because Snell's law uses the refractive index which is wavelength-dependent, it is then interesting to assign additional attributes to each ray, such as the wavelength. This will permit us to calculate the refractive index that should be used for each ray. The characteristics of each ray can also be extended to include the electric field amplitude. This allows extending the pure geometrical tracing to include physical models that take into account the optical element reflectivity or transmission. This combination of using a geometrical model plus a physical model allows to simulate every element used in a synchrotron beamline, such as mirrors and combinations of them (e.g., KBs); lenses, compound refractive lenses, and transfocators; gratings of any type and shape, crystal systems, etc. Ray-tracing is a simple and extremely powerful technique for calculating the main characteristics of the photon beam (size, divergence, photon distribution) at every point of the beamline. 
Several packages created in the synchrotron community and are available, like SHADOW \cite{codeSHADOW}, RAY \cite{codeRAY}, McXtrace \cite{codeMCXTRACE} or XRT \cite{codeXRT}. For this study, we use SHADOW and its interface ShadowOUI \cite{codeSHADOWOUI}, available as a module of the OASYS suite.

\begin{figure}
\label{fig:ray-tracing}
\centering
~~~~~ESRF-1~~~~~~~~~~~~~~~~~~~~~~~~~~~~~~~~~~~EBS
\newline
\includegraphics[width=0.95\textwidth]{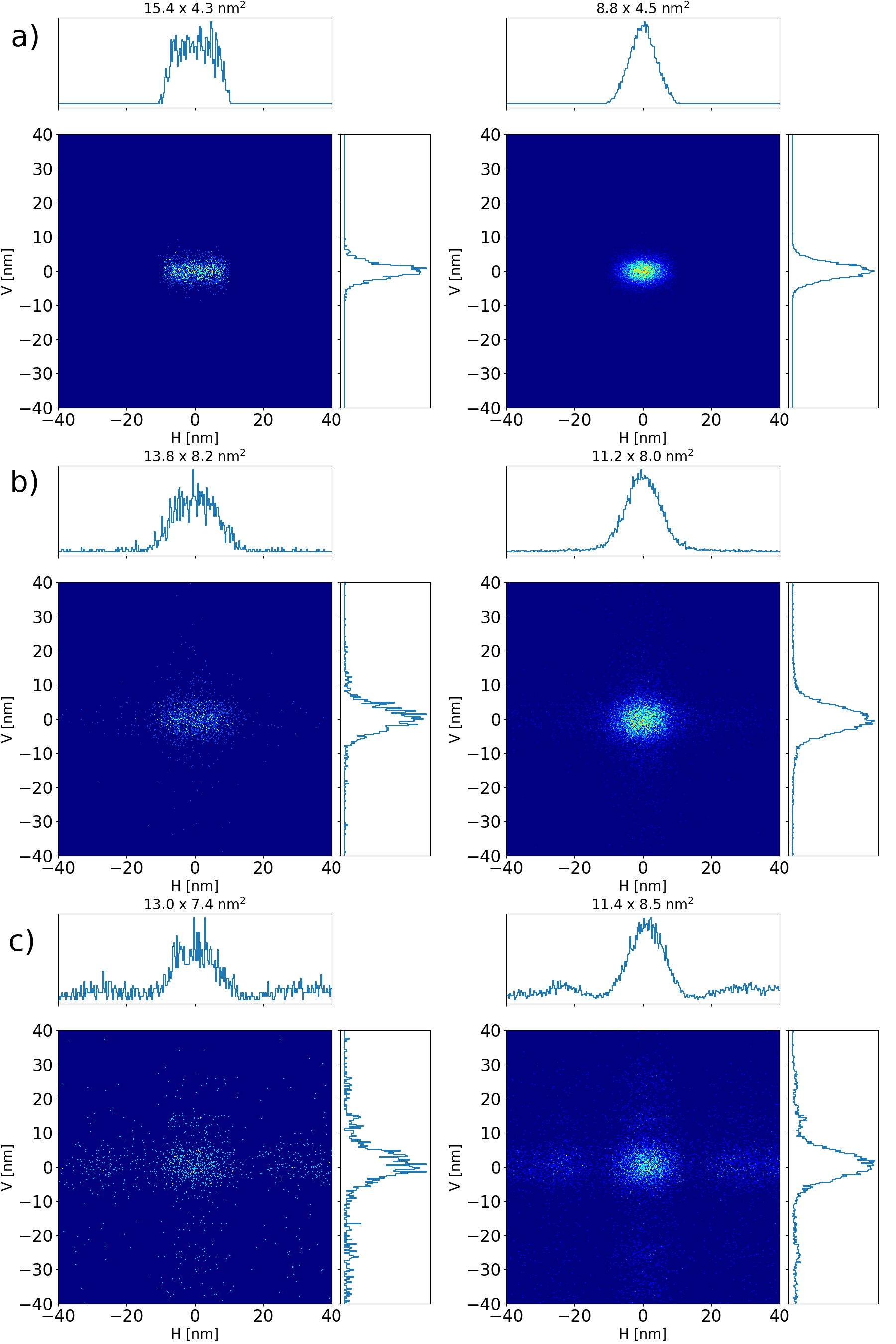}
\caption{Image produced by different types of ideal focusing for the ESRF (left column) and EBS (right column). The source follows Gaussian distributions with parameters in Table~\ref{tab:HandCalculations}. 
a) Simulations using the ideal reflectors (cylindrical for ML, elliptical for the BK mirrors) without slope errors. 
b) Images calculated using the hybrid method (ray-tracing plus coherent diffraction) with ideal reflectors with no slope errors. It shows the focal spot degradation due to the diffraction arising from the cropping of the beam by KB mirrors. c) Simulations using the hybrid method including slope errors from the real mirror profiles measured at the ESRF metrology lab (see Appendix~\ref{appendix:metrology}). It can be observed further beam broadening and the growth of satellite structures. 
The focal beam dimensions (FWHM of the histograms) are written in the plot title.
}
\end{figure}

We performed ray-tracing of the ID16A beamline using an undulator source tuned to have its resonance at 17225~eV and approximated using Gaussian distributions (Eqs.~\ref{eq:photon big sigmas}), considering the ML as a cylindrical mirror, and using elliptical cylinder reflectors with finite dimensions as the KB focusing elements. The KB reflector shape are elliptical as required for a point-to-point focusing. For a first calculation, the reflectors (ML and KBs) are considered perfect and have no slope errors. 
The results in Fig.~\ref{fig:ray-tracing}a show two phenomena. One is that the migration from ESRF-1 lattice to the EBS will significantly improve the horizontal size, and the VSS is not needed. For EBS, the whole beam passes though the VSS, producing a final spot with almost Gaussian horizontal intensity profile instead of the step distribution for ESRF-1. Secondly, one can observe a much larger luminosity in the EBS case. The transmission values are 0.16\% for ESRF-1 and 2.22\% for EBS. These values confirm the analytical estimations (0.2\% for ESRF-1 and 2.95\% for EBS). As mentioned before, these values are similar to the coherent fraction \cite{arxivCF} (0.12 for ESRF-1 and 2.73\% for EBS), meaning that a very high coherence of the beam is expected. It is therefore expected some broadening of the beam due to diffraction effects originated by the mirrors acceptance. 

The hybrid method \cite{hybrid} uses geometric ray-tracing combined with wavefront propagation. The code computes diffraction effects when the beam is cropped and can also simulate the effect of mirror figure errors when diffraction is present. The method can be applied to an entire beamline by simulating each optical element iteratively under the hypothesis of the validity of far-field propagation in all intermediate propagations. If this is not the case, a near field version is available for investigating imaging by individual optics. The interest of the hybrid method is to assess in a fast way the influence of beam coherence because the code is considerably faster than the full methods for dealing with the partial coherence of the synchrotron beam that are described below. Fig.~\ref{fig:ray-tracing}b presents the results using the hybrid model for the beamline with ideal elliptical optical surfaces. When compared with the pure ray-tracing in Fig.~\ref{fig:ray-tracing}a it can be appreciated that the spot is broadened. This is more evident in the vertical direction, because of the smaller focal size. The spot size is about 7-8~nm in V for both storage ring lattices, and in horizontal is about 15~nm for ESRF-1 High$\beta$ and 11~nm for EBS. It is interesting to note the almost round spot that will be obtained with the new EBS lattice. 

The surface figure of the optical elements is another limiting effect for determining the size and intensity of the beam at the image position. It is possible to infer the effect of mirror imperfections and in particular slope errors by only knowing some statistical parameters such as the RMS slope error and then building synthetic surface profiles to be used in the simulations. It is, though, highly recommended to use metrology data from real mirrors, when possible. When not available, one can use data from existing mirrors from a database (e.g., \cite{dabam}) and extrapolate some of the effects to the hypothetical new mirrors. This is the typical approach used when designing a new beamline where the mirrors do not exist yet. However, in our case, the mirrors already exist and have been extensively characterized at the ESRF optical metrology laboratory.  The metrology data of the mirrors for the beamline under study are presented in Appendix~\ref{appendix:metrology}. These experimental data are used in all the simulations in this paper. Simulations using the hybrid method including slope errors are shown in Fig.~\ref{fig:ray-tracing}c. A small broadening of the peak due to slope errors can be observed, but, more importantly, new satellite structures close to the main peak appear.

%

\subsection{Wave optics simulations}
\label{level2}

In this section the coherence properties of the beamline are analyzed using, again, different approaches distinguished by their complexity and necessary computer resources. We first analyze the beamline supposing a fully coherent beam. Subsection~\ref{wofry} shows a simplified 1D model. Subsection~\ref{srw_se} uses a full 2D model with accurate simulation of the undulator emission but without considering storage ring emittance. These simplifications allow the estimation of diffraction effects by apertures and slope or figure errors. Subsequently, we introduce methods to treat the partial coherence originated by the storage ring emittance. Two variants are shown, a Monte Carlo approach (subsection~\ref{srw_me}) that permits us to calculate in great detail the intensity distribution of the focused beam, and a decomposition of the radiation in coherent modes (subsection~\ref{comsyl}) that allows quantitative evaluation of the coherent beam characteristics.

\subsubsection{Simplified wave optics simulation (coherent case in 1D with point source and ideal elements)}
\label{wofry}

\begin{figure}
\label{wofry1D}
\centering
\includegraphics[width=0.95\textwidth]{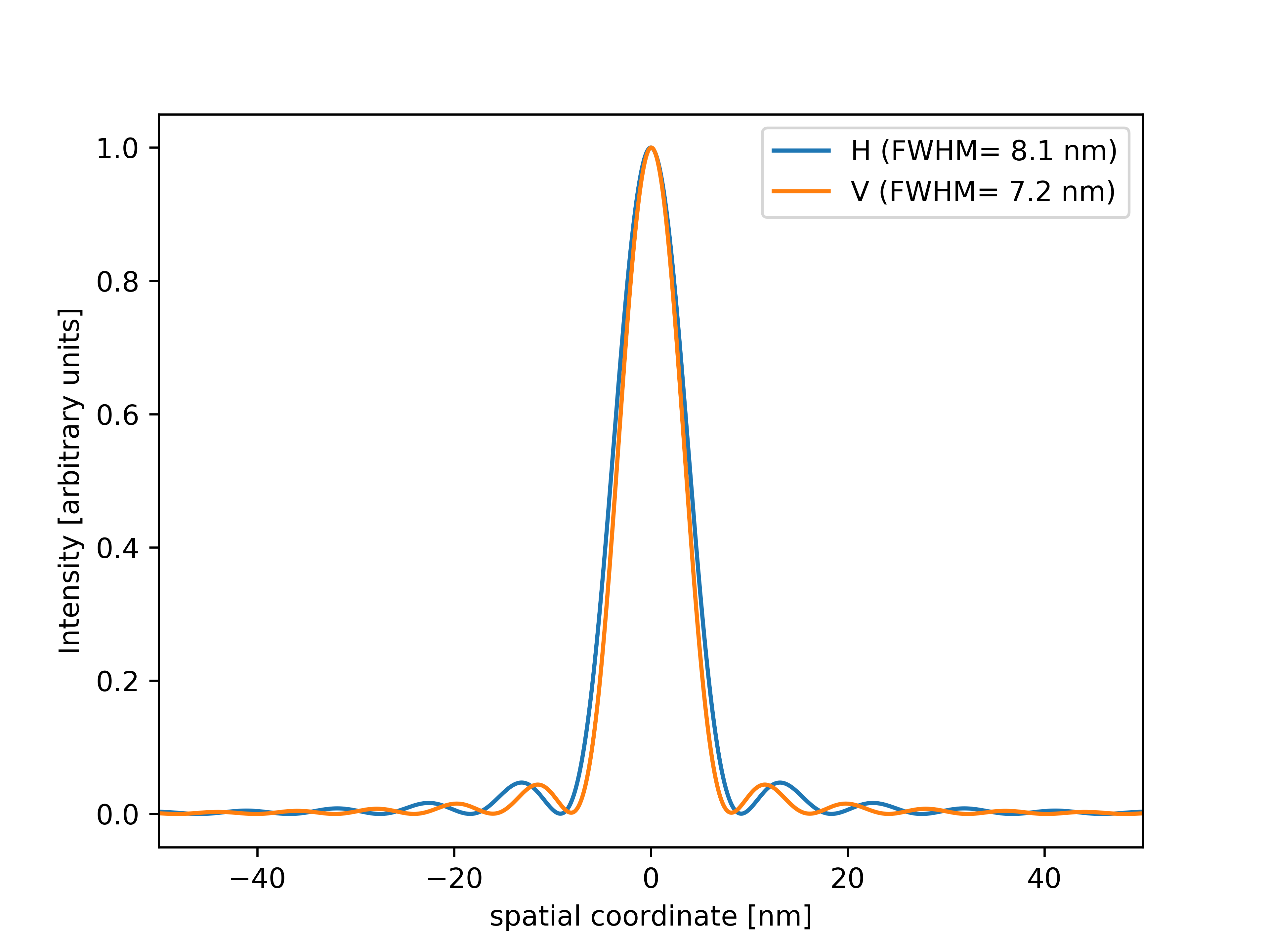}
\caption{Intensity distributions at the focal position (horizontal and vertical cuts) as calculated by a simplified 1D waveoptics model. 
}
\end{figure}

As discussed in Section~\ref{level0}, the clipping of the beam, mainly due to the acceptance of the KB mirrors produces a significant broadening of the focal spot.  Very simple calculation of the diffraction limit can be done using a convergent spherical beam clipped by a slit of 400$\times$400$~\mu$m$^2$ placed 5~cm upstream of the focus (at the second KB mirror position), producing a spot of about 7.75$\times$7.75~nm$^2$, therefore qualifying this beamline to be a diffraction-limited beamline. We perform here a simplified wavefront simulation in 1D, thus decoupling the simulations of the horizontal and vertical planes. The source can be simplified by using a point source (spherical wavefront) and applying a Gaussian intensity profile with $\sigma$=$\sigma'_u d$ with $\sigma'_u$ given by Eq.~\ref{eq:photon small sigmas} and $d$ the distance from the point source to the observation plane. Each optical element can be simplified by separating its action into two parts: the focusing behavior implemented as an ideal lens, and the finite size because of the boundaries, implemented as a slit. The WOFRY package \cite{codeWOFRY} in OASYS was used for these simulations. The beam size at different positions of the beamline is shown in Table~\ref{tablewofry1D}. The final image profiles are in Fig.~\ref{wofry1D}.

\begin{table}[]
    \centering
    \caption{Beam sizes calculated with simplified wave optics propagation (point source and ideal lenses with aperture for modeling the element dimensions). The zoom or scaling factor used for the propagation is also displayed. 
    }
    \label{tablewofry1D}
    \resizebox{\textwidth}{!}{%
    \begin{tabular}{cccccc} 
    Element:                & ML            & VSS          & KBv               & KBh           & sample        \\
    Position [m]:           & 28.3          & 40.00        & 184.90            & 184.95        & 185.00        \\
    H beam size [$\mu$m]:   & 329           & 1.11         & 4084              & 390           & 0.008         \\
    V beam size [$\mu$m]:   & 329           & 466          & 899               & 450           & 0.007         \\ 
    H scaling factor:       & 1.0           & 0.01         & 440               & 1             & 0.00002       \\
    V scaling factor:       & 1.0           & 0.64         & 5                 & 0.5           & 0.00002       \\
    \end{tabular}%
    }
\end{table}

\subsubsection{Full wave optics simulation for coherent case}
\label{srw_se}

A next step in the wave optics simulation is, assuming full coherence, to generate a 2D source wavefront that matches in detail the real source (an undulator in a storage ring) and then simulate the propagation of such a wavefront through the optical elements with models that reproduce well the characteristics of the real elements. Once the source is defined, an approach of how best to simulate the beamline optical elements. 
Once the source is defined, several approaches of increasing complexity are used: a) use ideal thin lenses preceded by slits to emulate finite apertures (as done in the 1D model presented in Section~\ref{wofry}) b) use grazing incidence focusing optics models \cite{Canestrari2014}; and c) introduce the effect of imperfect optics on the simulations, i.e. slope and figure errors to both ideal lenses and mirrors. The computer code chosen for these calculations is ``Synchrotron Radiation Workshop'' (SRW) \cite{codeSRW}, which over the past two decades has been extensively benchmarked throughout different synchrotron radiation facilities and has become widely accepted within the X-ray community. 



\begin{figure}
    \centering
        \includegraphics[width=0.95\textwidth]{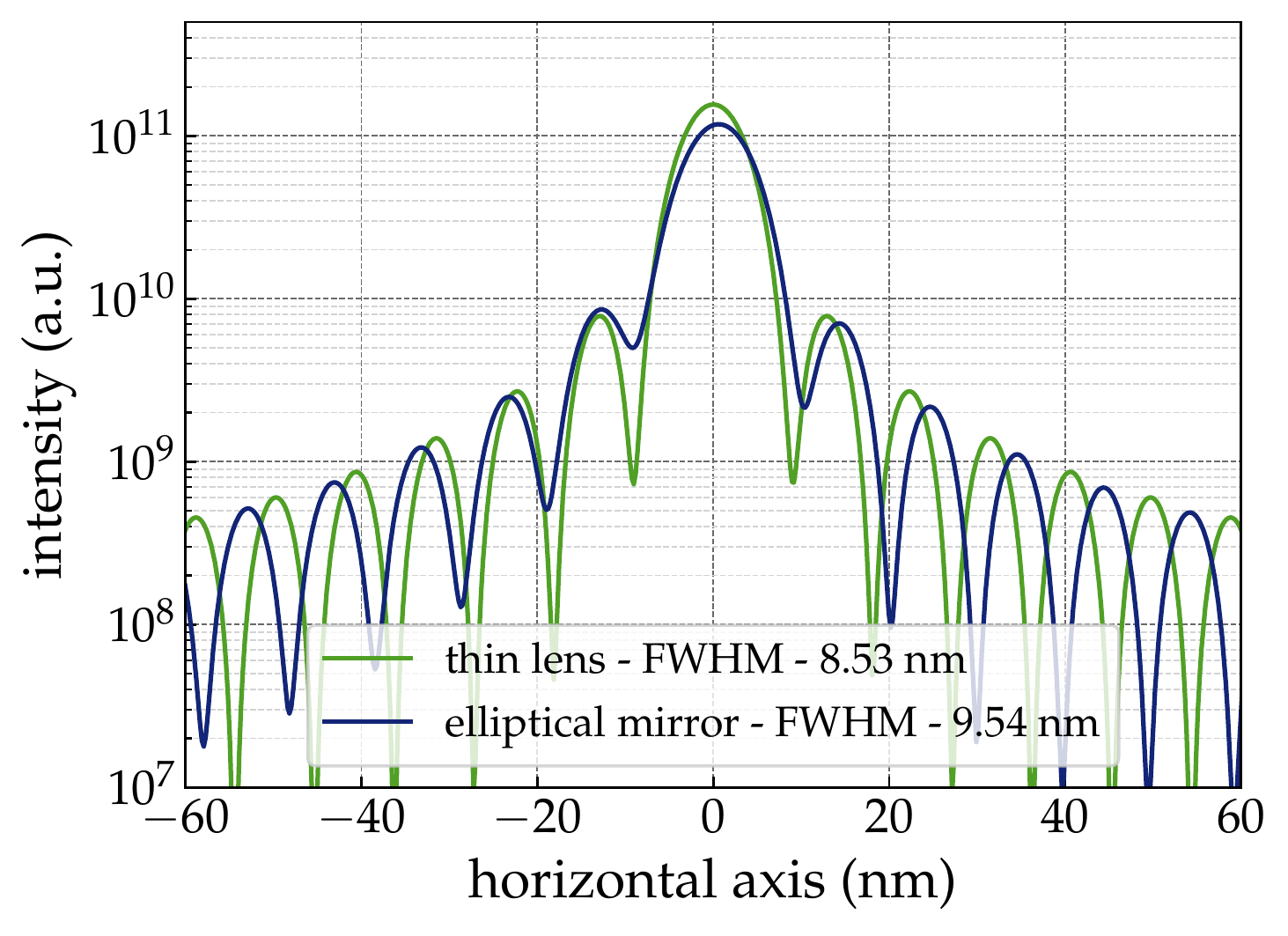}
        \includegraphics[width=0.95\textwidth]{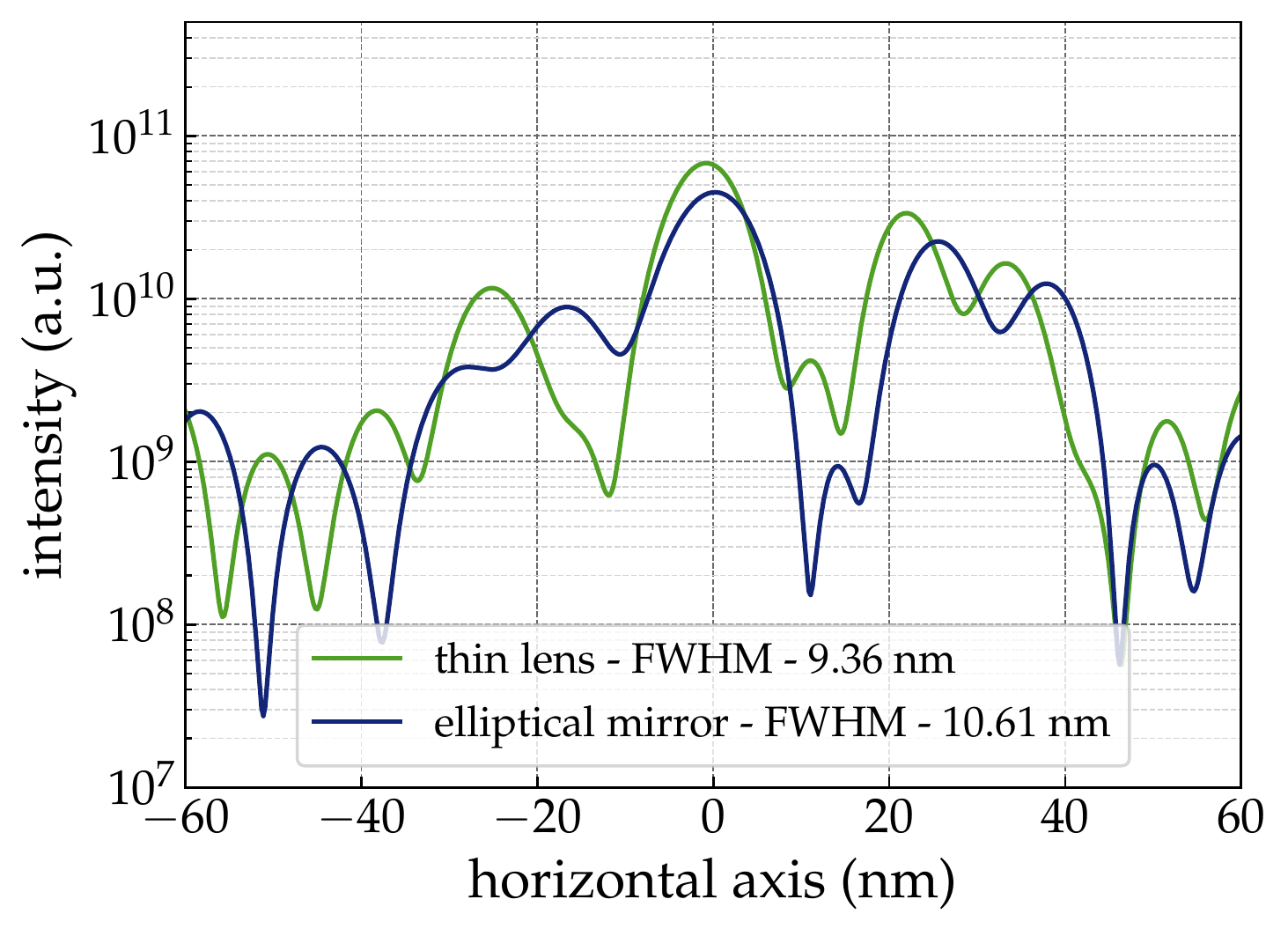}
    \label{fig:SingleElectron}
    \caption{Intensity profile at the sample position produced by the single-electron emission propagated through the ID16A beamline. Top: the perfect focusing with ideal lenses (green) is compared with the focusing using elliptical mirrors (blue). Bottom: the same as in the previous plot, but adding the effect of the mirrors slope errors in both cases.}
\end{figure}

Figure \ref{fig:SingleElectron} shows the horizontal intensity profile for a filament beam (no emittances). It can be seen that there is little difference between using thin ideal lenses or the elliptical mirror model. The elliptical mirror presents a slight reduction in intensity and asymmetry but no significant differences are observed, which is in agreement with the findings of \cite{Canestrari2014}. Notice that the differences are minimal and to accentuate them we have chosen logarithmically scaled plots to represent the intensity distributions at the focal position. It is also important to mention that this equivalency between ideal lenses and elliptical mirrors is only valid close to the focal position: down- or upstream the focal position, differences are no longer negligible - c.f. Fig.~5 in \cite{Canestrari2014}.  Increasing the complexity of the model increments considerably the simulation time, which went from typically $\sim16$~s to $\sim24$~s. 


It is also possible to add slope errors to the simulations to study the resulting impact upon the focal spot quality (Fig.~\ref{fig:SingleElectron}b). For that, the available surface metrology data of the reflective elements (see Appendix~\ref{appendix:metrology})  are used. For both cases (i.e., simulating the beamline using thin optical elements or elliptical mirrors), the beam sizes calculated as the FWHM at the focal position are little affected by the presence of the slope errors (see Table \ref{tab:SRW_results}). There is, however, a significant reduction and blurring of the interference fringes compared to observations with no slope error. They are almost completely removed and associated with a large increase of the background level: part of the radiation in the main peak shifts to the side lobes when the slope errors are considered. This is accompanied by an obvious reduction in the peak intensity.  Adding the mirrors figure errors to the model slightly increases the simulation time: a simulation of a filament beam using elliptical mirrors with figure errors runs in less than 30s. 

\begin{table}\label{tab:SRW_results}
\centering
\caption{Calculated beam dimensions (FWHM, in nm) at the focal position for different models of focusing elements using SRW.}
\resizebox{\textwidth}{!}{%
\begin{tabular}{ccccc}
Focusing elements                 & Plane      &  Single electron & EBS         & ESRF-1       \\ \\
ideal lenses/ elliptical mirrors  & Horizontal &  8.53/9.54       & 12.71/11.56 & 17.75/12.41 \\
the same, with errors             & Horizontal &  9.36/10.61      & 13.20/12.52 & 17.78/15.71 \\
ideal lenses/ elliptical mirrors  & Vertical   &  7.22/7.94       & 8.66/9.38   & 8.90/10.11    \\
the same, with errors             & Vertical   &  7.70/8.90       & 8.42/9.14   & 8.90/9.86    \\                                            
\end{tabular}
}
\end{table}

\subsubsection{Full wave optics simulation: partial coherent case by means of multi-electron Monte Carlo sampling}
\label{srw_me}

The wavefront propagation methodology is based on the sampling of a single wavefront (monochromatic and coherent) in a 2D space (in horizontal and vertical spatial dimensions) at a given point of the beamline. This initial point considered as source cannot be the source physical position, because i) a point or spherical wave would concentrate all its intensity in a single pixel so the wavefront is ill-defined, and ii) the theory (e.g., \cite{jackson}) used for calculating of  synchrotron emission (i.e., undulators) permits the calculation of the radiation in positions far from the electron trajectory. The limitation of the wavefront propagation method is that a single wavefront is not representative for a beam that: in storage rings it is produced by the incoherent supperposition of the emission of many electrons circulating through the corresponding magnet structure. 

Diffraction limited sources would emit radiation that can be described in good approximation by a single wavefront. But even for new ultra low-emittance storage rings (improperly called diffraction-limited sources, see discussion in \cite{arxivCF}) the contribution of the electron dimensions ($\sigma_{e(h,v)}$ and $\sigma'_{e(h,v)}$) is not negligible as compared with the natural emission of radiation ($\sigma_u$ and $\sigma'_u$ for undulators as in Eq.~\ref{eq:photon small sigmas}). To account for this effect one should complete or extend the wavefront simulations to take into account that in fact, not only a single wavefront contributes to the properties of the beam. 

\begin{figure}
    \centering
        \includegraphics[width=0.95\textwidth]{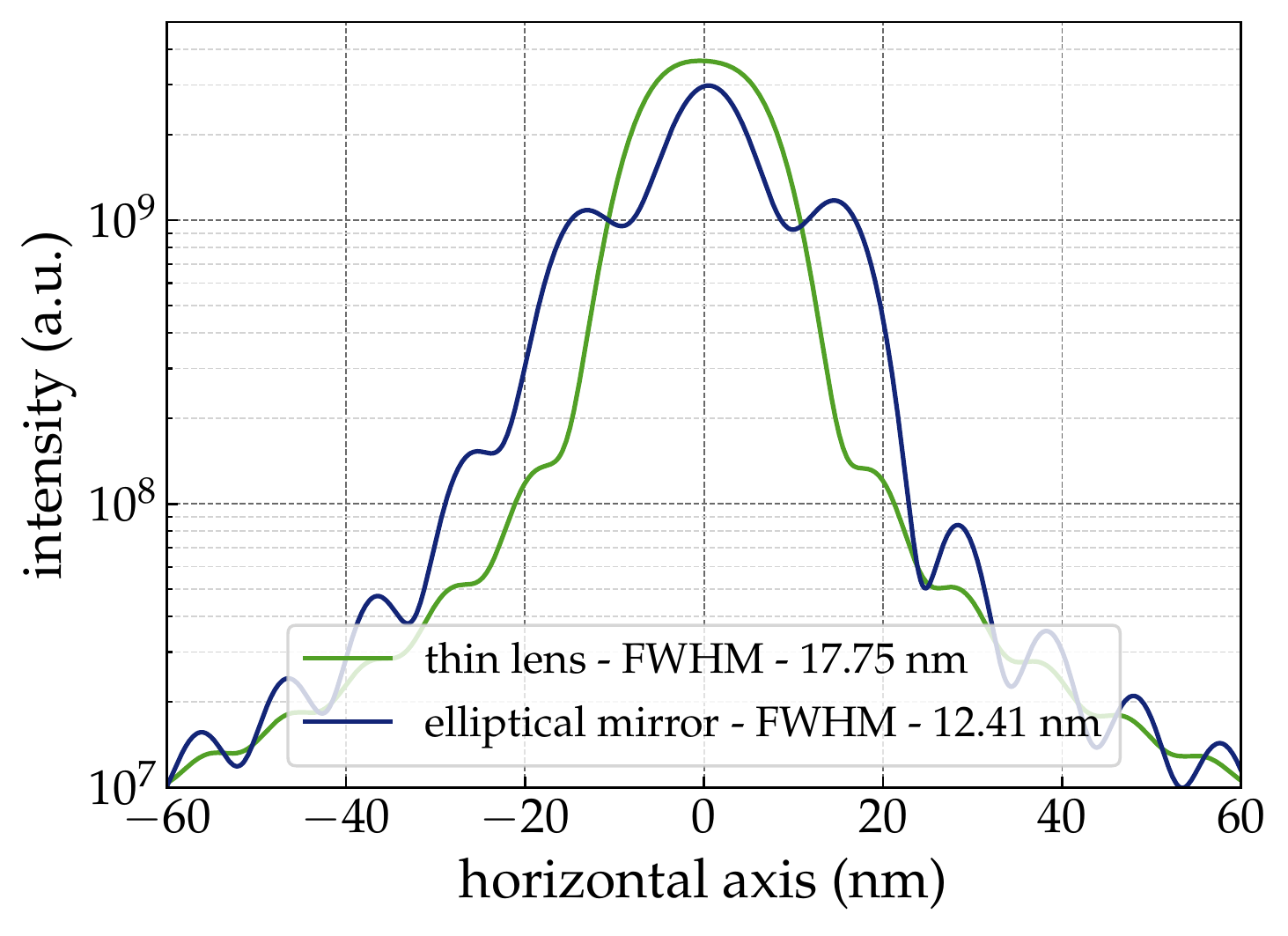}
        \includegraphics[width=0.95\textwidth]{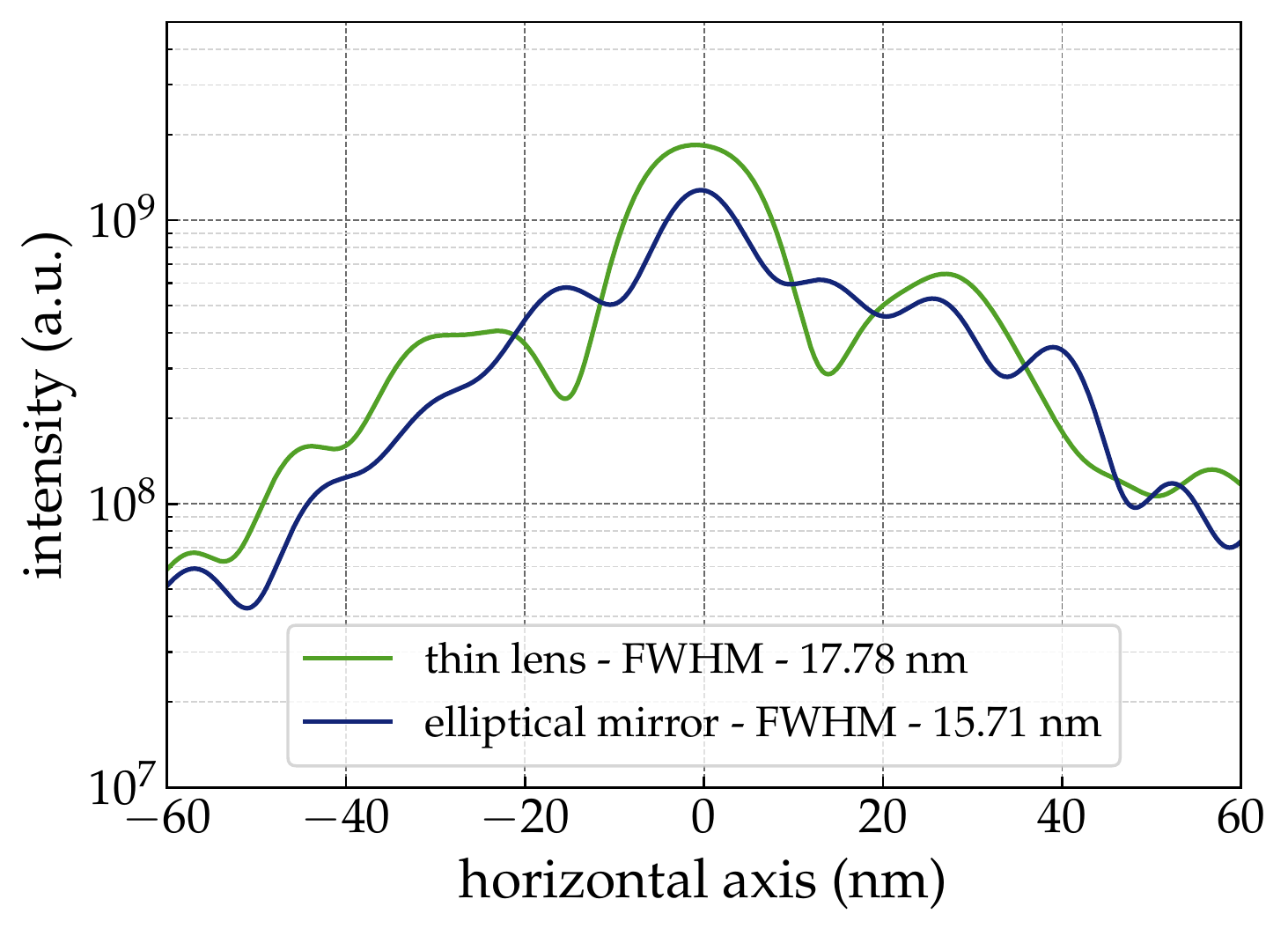}
    \label{fig:FiniteEmittanceA}
    \caption{
    Intensity profile at the sample position produced by multi-electron method sampled with parameters of the ESRF-1 lattice, using ideal focusing elements (top), and including the effect of the mirrors slope errors (bottom).
    }
\end{figure}

There are three approaches currently used for including the electron emittance effects in the wave optics simulations: i) use of convolutions, ii) Monte Carlo multi-electron sampling, and iii) Propagation of coherent modes. The convolution method consists of calculating the emission characteristics and then convolving the resulting intensity with Gaussians describing the electron beam. This method is simple and fast, as only requires simulation of a single wavefront. However, it presents two important limitations: it can only be applied to the intensity calculations and not to other parameters of interest for describing coherence, such as the degree of coherence, etc.; and it can only be applied at a plane right after the source position (it is usually unknown how the different optical elements will affect the source contribution to the phase space). The next method, Monte Carlo multi-electron propagation, is a statistical method based on the idea that one can exactly calculate the emission and its propagation for one electron that is deterministically described by its initial conditions at the entrance of the magnetic structure. It is then possible to perform many simulations for different initial conditions of the traveling electron and make the propagation along the beamline until the observation plane. Then, the different wavefronts (one for each electron) are combined to obtain a statistical map of the intensity and other magnitudes useful to describe partially coherent beams. This method presents two difficulties. The first one is convergence: it is not possible to know a priori how many electrons must be sampled to obtain a given level of accuracy. The second is computing resources. The complex task of calculating the emission of one electron, propagated it through every element of the beamline socoring the results must be repeated thousand of times. The third method, the propagation of coherent modes, will be discussed in the next section. 

\begin{figure}
    \centering
        \includegraphics[width=0.95\textwidth]{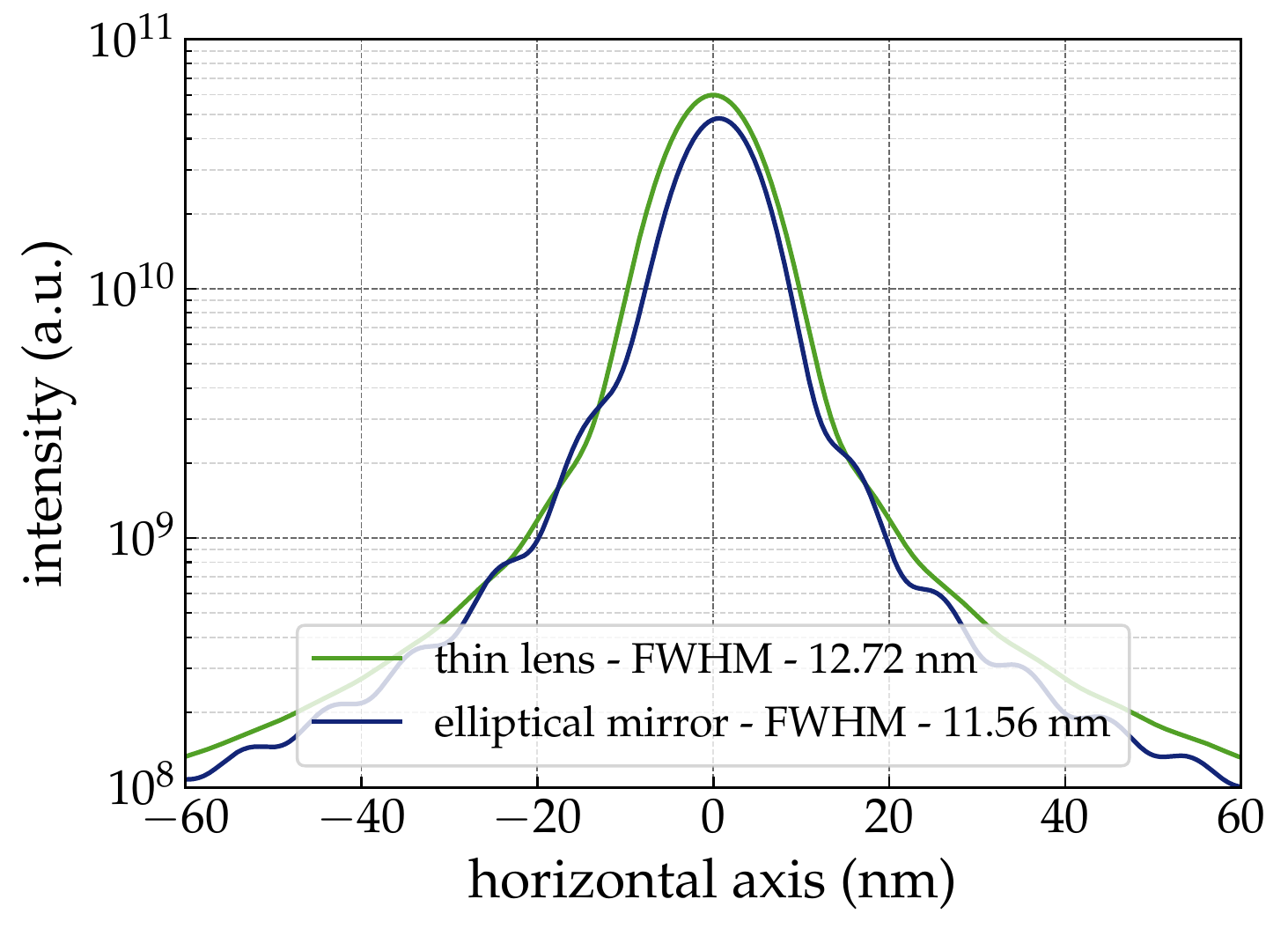}
        \includegraphics[width=0.95\textwidth]{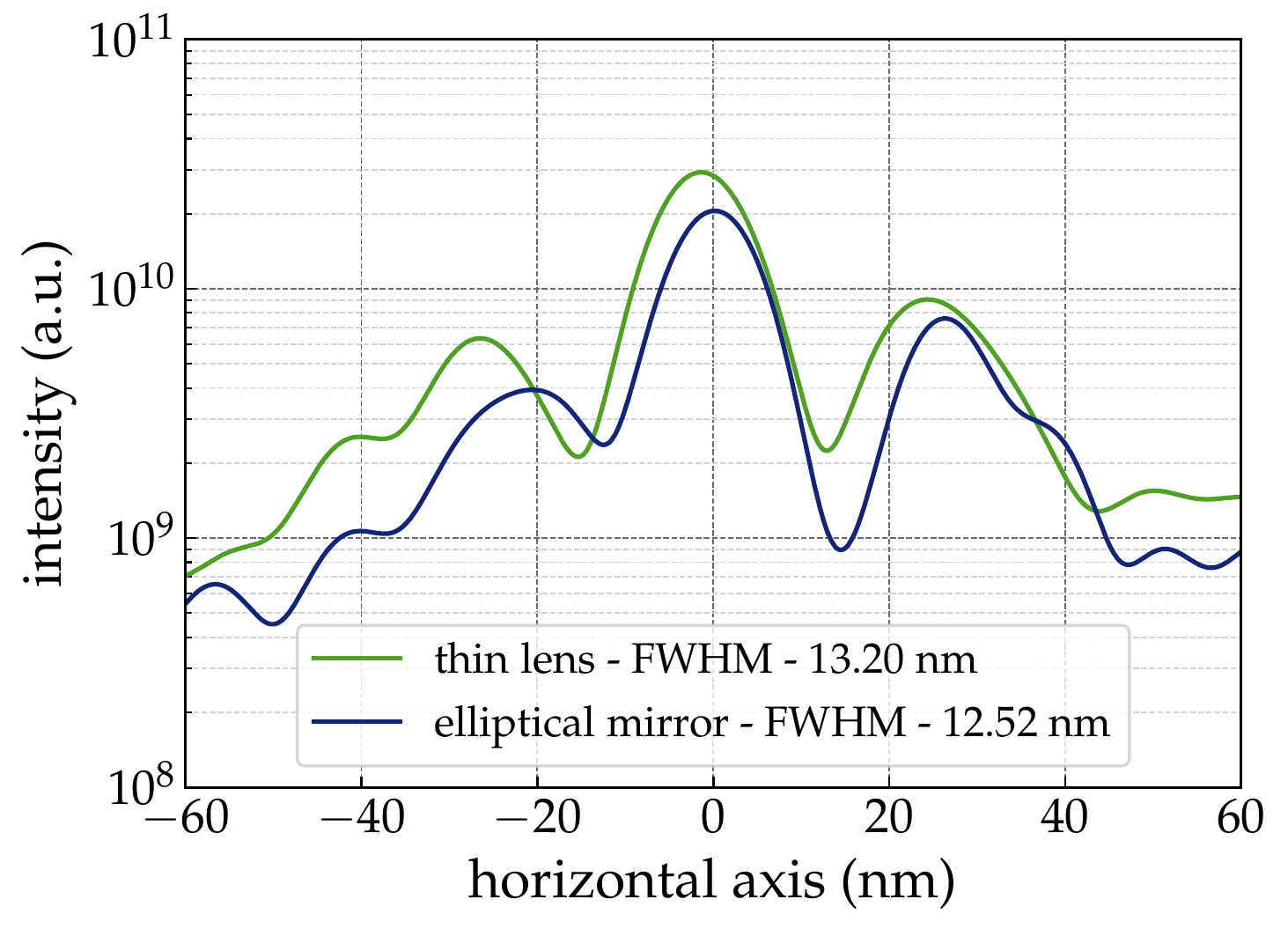}
    \label{fig:FiniteEmittanceB}
    \caption{
    Intensity profile at the sample position produced by multi-electron method sampled with parameters of the EBS lattice, using ideal focusing elements (top), and including the effect of the mirrors slope errors (bottom).
    }
\end{figure}

The results presented in this section are calculated using the Monte Carlo multi-electron method. It has been proposed by O. Chubar and it is fully implemented in SRW \cite{Chubar2011b}. The simulations presented here are built upon the ones described in the previous section (\ref{srw_se}). The different initial conditions were sampled from the 6D-electron phase space: its position on the cartesian space $(x_e, y_e, z_e)$; its direction in both vertical and horizontal planes  $(x'_e, y'_e)$ and its energy (as the electron energy spread is taken into account). Although simulations for highly coherent beamline converge rapidly (a few electrons are sufficient), this is in no way the case of the ID16A at ESRF-1: the VSS and the KB aperture reject a very high number of photons. To ensure good statistics when averaging the intensities of the different wavefronts from the sampled electrons, one has to go to a very high number of iterations. For the simulations presented here, 50000 wavefronts were used, amounting to a $\sim7$~h simulation time on a 56-core computer cluster. Convergence is faster for the EBS lattice as the electron-beam phase space is smaller.  Results are presented in Figures \ref{fig:FiniteEmittanceA} and \ref{fig:FiniteEmittanceB} and Table \ref{tab:SRW_results}.

The most important feature of the partially-coherent simulations, when compared to one of a filament-beam, is that the refraction fringes from the KB system aperture are smoothed out, although still present. The trends shown for the zero-emittance case are also found here, such as the reduction of the peak intensity and asymmetries when comparing the simple lens model with the elliptical mirror (although here it is more subtle). The major difference found when comparing the ESRF-1 and the EBS is, once more, the beam intensity at the sample position, thus permitting the beamline after the upgrade to exploit a more brilliant beam, that is, more photons on the same phase space. 


%

%
%
%
\subsubsection{Full Wave optics simulation: partial coherence case by means of coherent mode decomposition}
\label{comsyl}
%

The cross-spectral density (CSD) (sometimes called mutual intensity) completely describes the coherence of a beam. 
At a given plane at a distance $z$ from the source, for a photon beam of frequency $\omega$, the CSD is the correlation of the radiated electric fields between two spatial points $(x_1,y_1)$ and $(x_2,y_2)$:

\begin{equation}\label{CSD}
W(x_1,y_1,x_2,y_2;z,\omega) = <E^{*}(x_1,y_1; z,\omega) E(x_2,y_2;z,\omega)>
\end{equation}
where $E$ is the amplitude of the electric field, and $<>$ stands for the ensemble average. The CSD is at the origin of any other information about the coherence of the beam, like the spectral density $S$ (or, spatial distribution of the intensity): 
\begin{equation}\label{SD}
S(x,y,;z,\omega) = W(x,y,x,y,;z,\omega)
\end{equation}
or the spectral degree of coherence $\mu$:
\begin{equation}\label{SDC}
\mu(x_1,y_1,x_2,y_2;z,\omega) = \frac{W(x_1,y_1,x_2,y_2;z,\omega)}{\sqrt{ S(x_1,y_1;z,\omega) S(x_3,y_2;z,\omega}}
\end{equation}
that measures the coherence of the beam between two points, and ranges from fully incoherent ($\mu=0$) to fully coherent ($\mu$=1). 
Eq.~\ref{CSD} can in principle be calculated using Kim's convolution theorem \cite{kim1986} (see \cite{geloni2008} for a full discussion on its validity) but its storage is prohibitive in present computers: supposing sampling one spatial dimension by $N\approx1000$ pixels, a wavefront will contain $N^2$ pixels, and the CSD will be sampled over $N^4$ points. They are complex numbers (16 bits) so the only storage is of the order of GB or TB. Considering also that the propagation of $W$ from a beamline position $z_1$ to another position $z_2$ will need 10$^8$ to 10$^{12}$ 4D integrals, one can easily imagine the physical impossibility to perform such calculations. 

A solution to this problem is the expansion of the CSD in coherent modes \cite{mandel_wolf}:  
\begin{equation}\label{decomposition}
 W(x_1,y_1,x_2,y_2;z,\omega) = \sum\limits_{m=0}^{\infty} \lambda_m \Phi_m^{*}(x_1,y_1;z,\omega) \Phi_m(x_2,y_2;z,\omega)
 \end{equation}
where $\lambda_m$ are the eigenvalues or weights, and $\Phi_m$ are the eigenfunctions or (orthonormal) coherent modes. A theoretical study and a numerical algorithm for the coherent mode decomposition of undulator emission in storage rings has been recently proposed  \cite{GlassThesis,GlassEPL} and is implemented in the computer package COMSYL \cite{codeCOMSYL}. The innovative concept is the numerical description of the undulator CSD in terms of its coherent modes $\Phi_m(x,y;z,\omega)$ and eigenvalues $\lambda_m $. One can define the occupation of coherent modes as
\begin{equation}
 d_m = \frac{\lambda_m}{\sum\limits_{i=0}^{\infty} \lambda_i},
\end{equation}
or the cumulated occupation like
\begin{equation}
 c_m = \sum\limits_{i=0}^{m} d_i.
\end{equation}
An important property of the coherent mode decomposition is that it optimizes the representation of the CSD when the series is truncated. In other words, truncating the series in Eq.~\ref{decomposition} up to the term $m$, the CSD obtained approximates better the exact CSD (this with infinite terms) than any other expansion. It also implies that the first term is more important than the second, and so on. The first or lower term (index zero) is the most important coherent mode. It permits the definition of the coherent fraction (CF) as the occupation of the lower coherent mode: CF=$d_0$. Obviously, a more coherent source will require fewer modes than a mostly incoherent source to approximate the CSD to a given precision.

\begin{figure}\label{fig:histomodes}
    \centering
        \includegraphics[width=0.95\textwidth]{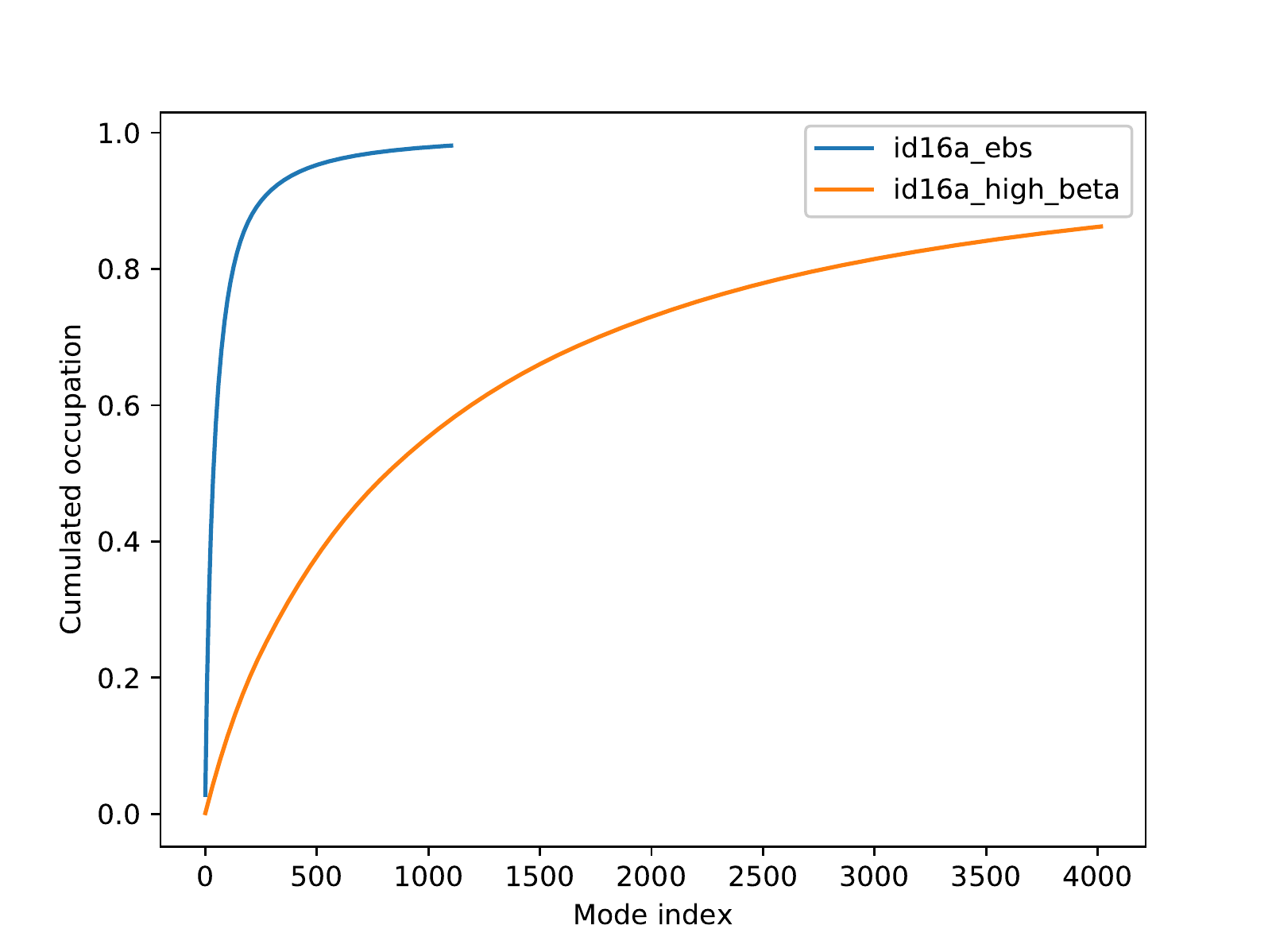}
        \includegraphics[width=0.9\textwidth]{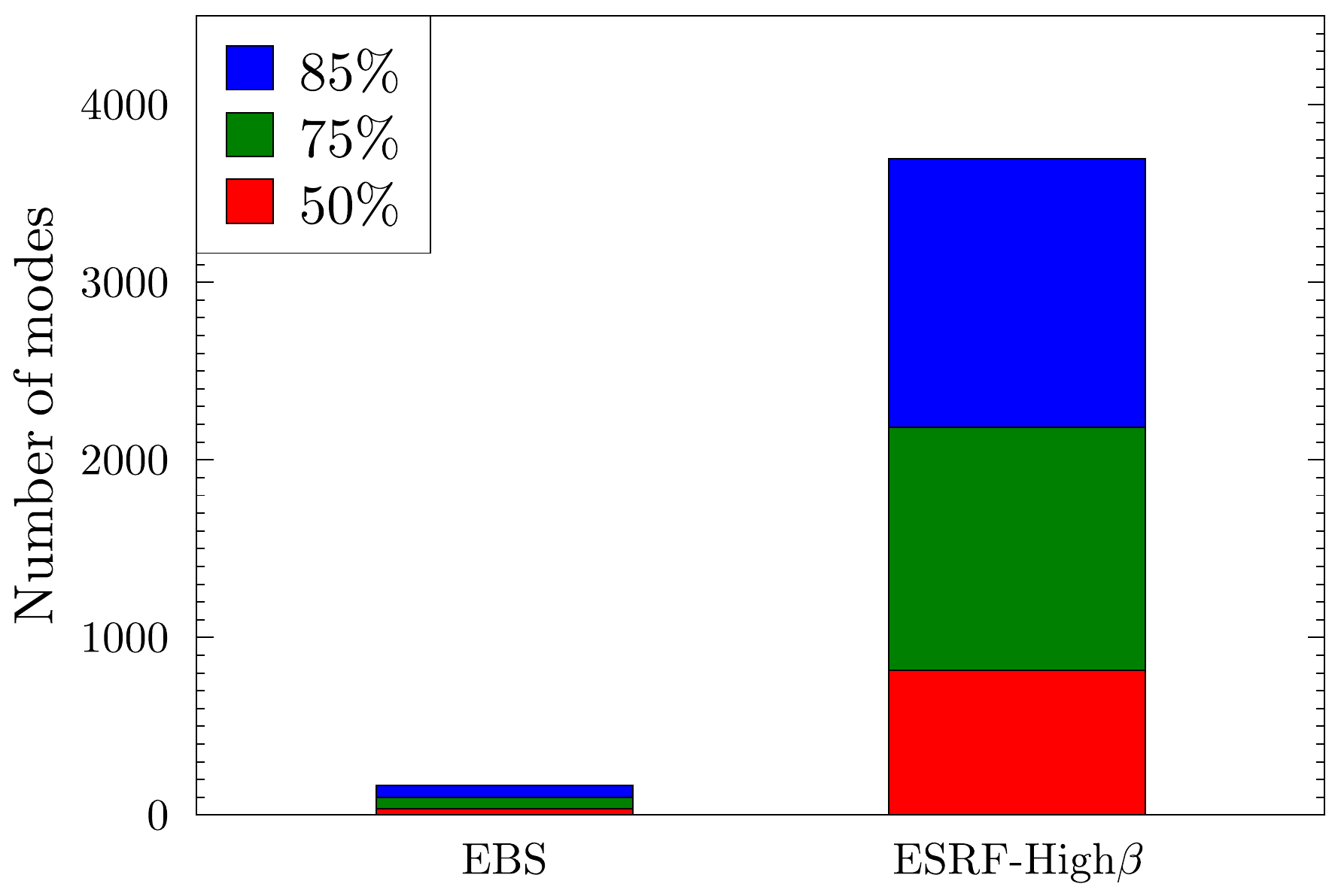}
    \caption{Cumulated occupation of the coherent modes for the EBS and ESRF-High-$\beta$ (top), and number of modes needed to reach a given percent of CSD occupation (bottom). }
\end{figure}

We performed COMSYL calculations for the U18.3 undulator radiation of ID16A of 1.4~m length using both the EBS and ESRF-1 lattices. The obtained CF is 2.8\% for EBS and 0.13\% for ESRF-1 High$\beta$, indicating that the overall coherent flux emitted by EBS will be about 20 times the existing coherent flux with ESRF-High$\beta$. Fig.~\ref{fig:histomodes}a shows the cumulated occupation of the coherent modes for the full central cone undulator emission. It can be noticed that the EBS cumulated occupation grows very steeply with the number of coherent modes, whereas for ESRF-1 the number of coherent modes needed to attain a given value of cumulated occupation much higher. This can also be shown in Fig.~\ref{fig:histomodes}b, where the number of modes to fill 50\%, 75\% and 95\% of spectral density is compared for EBS and ESRF-1.

\begin{figure}\label{fig:spectraldensity}
    \centering
        \includegraphics[width=\textwidth]{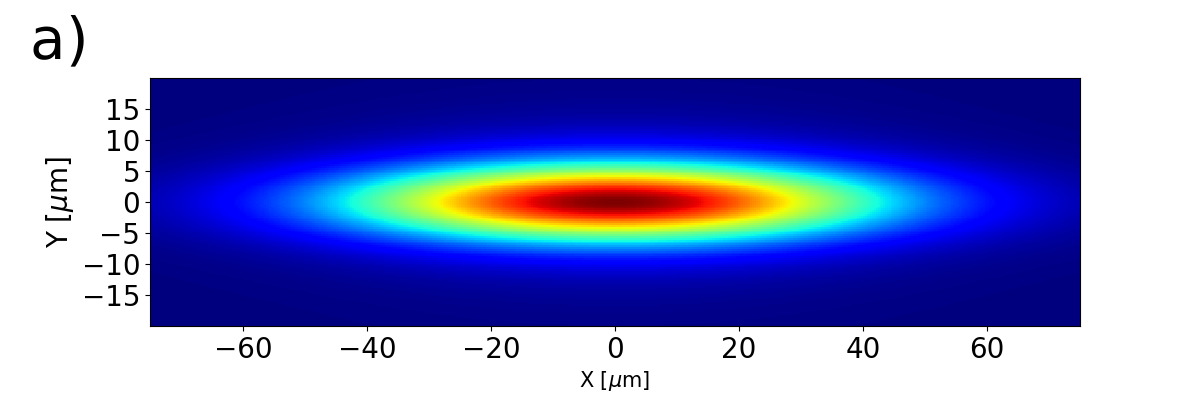}
        \includegraphics[width=\textwidth]{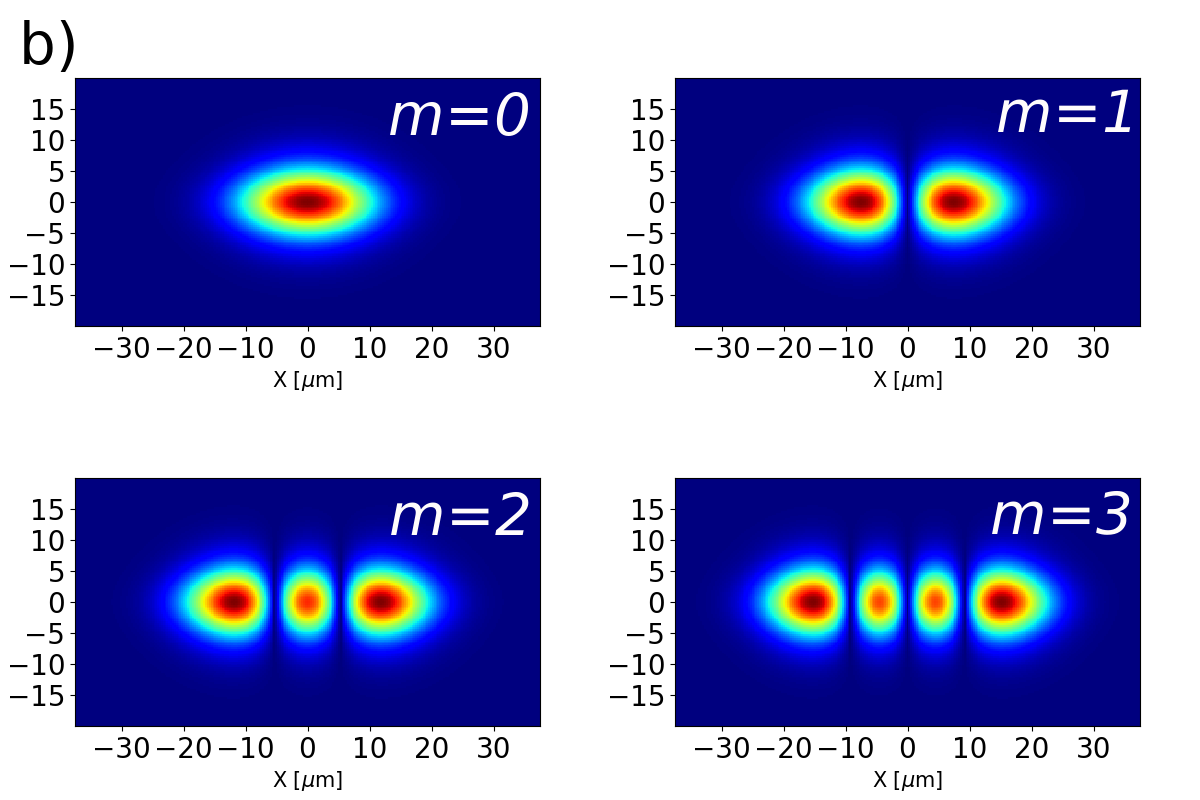}
    \caption{a) Spectral density $S(x,y)$ at the source position for EBS. b) Intensity distribution of the first coherent modes ($m$=0 to $m$=3) at the source position.
    }
\end{figure}

The spectral density at the EBS photon source is represented in Fig.~\ref{fig:spectraldensity}a. It shows an elliptical shape with a horizontal size larger than the vertical.  The spectral density is reproduced by the addition of the coherent modes weighted by the eigenvectors (from Eq.~\ref{decomposition})
\begin{equation}
S(x,y) = \sum_{m=0}^{\infty} \lambda_m |\Phi_m(x,y)|^2.
\end{equation}
Figsures~\ref{fig:spectraldensity}b-e represent the coherent modes with $m$=0 to $m$=3. It can be shown that the first mode is centered on the axis but extends over a spatial region smaller than the spectral density. The next mode ($m$=1) contributes to extend the spectral density in horizontal but presents zero intensity at the origin. Successive modes extend the spectral density in horizontal, sometimes presenting a lobe center at zero and sometimes not (e.g., $m$=3). Modes with several lobes along the vertical will also appear, but not so often than in the horizontal, because of the elliptical shape of the spectral density. The coherent modes are orthonormal, thus verifying 
\begin{equation}\label{eq:mode normalization}
M_{mn} = \int_{-\infty}^{\infty} \Phi^*_m(x,y) \Phi_n(x,y)~dx~dy = \delta_{mn}
\end{equation}
with $\delta_{mn}$ the Kronecker delta. The coherence fraction of 2.8\% for the EBS source means that for an optimal coherence experiment one should remove 97.2\% of the emitted photons. This task is performed by the beamline. The beamline elements modify the different modes in a different manner, but it will never be possible to remove all the photons except from the lowest coherent mode because there is an overlapping of the intensity of the lowest mode with the other ones. Therefore it is impossible to obtain a fully coherent beam containing only a single mode. 

\begin{figure}\label{fig:mode transmission}
    \centering
        \includegraphics[width=\textwidth]{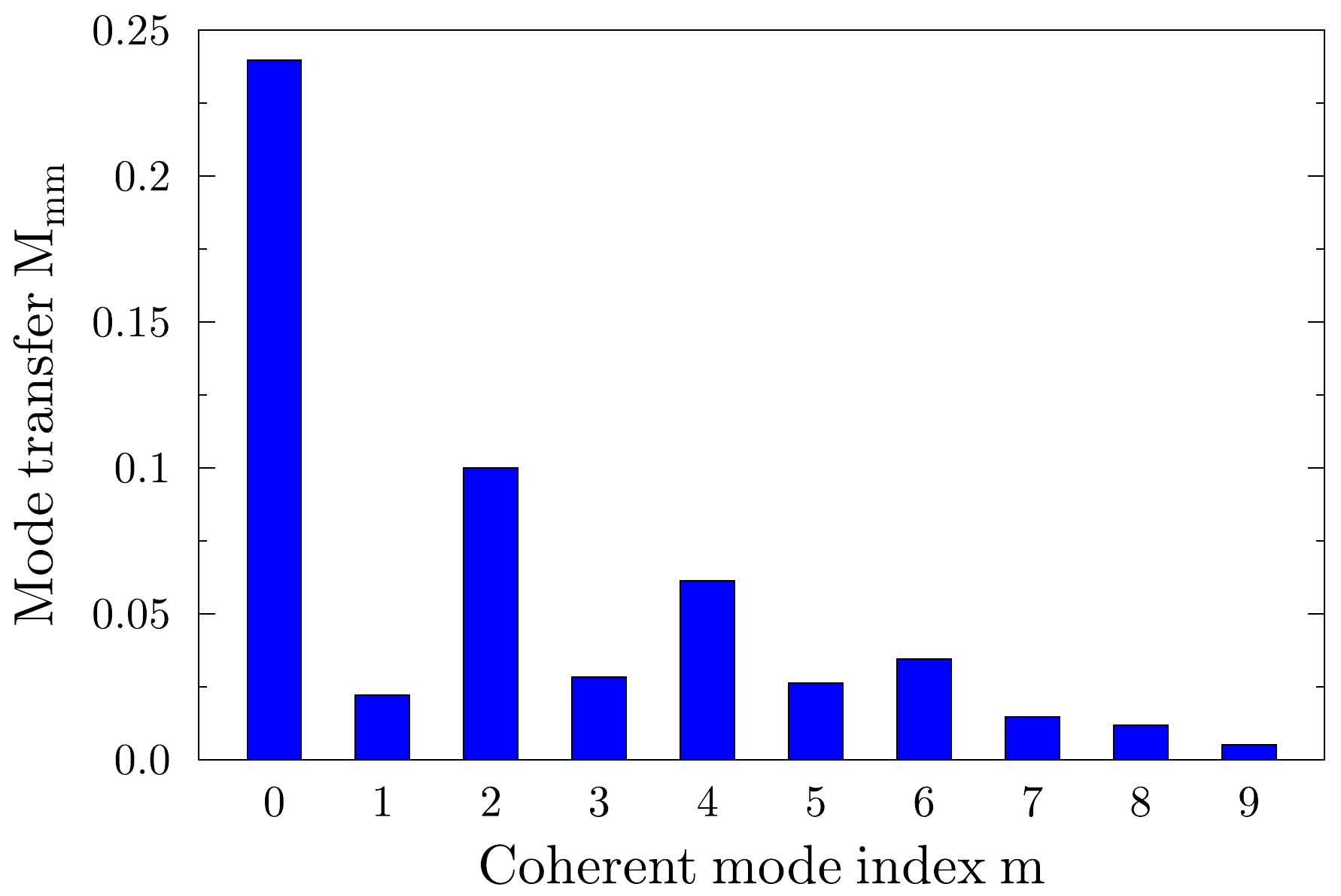}
    \caption{Transmission $M_{mm}$ (see Eq.~\ref{eq:mode normalization}) of the different modes of the EBS source  due to the effect of propagation and cropping by the beamline optics.}
\end{figure}

The coherent modes calculated by COMSYL at the EBS source have been propagated in OASYS using the WOFRY package implementing a simplified beamline with ideal focusing elements and slits to determine their numerical aperture (like in Section~\ref{wofry}). Fig.~\ref{fig:mode transmission} shows how the intensity of the first modes is reduced by geometrical considerations. 
As discussed, coherent modes will be ``cut'' in a different manner, depending on the effect of the slit.
This implies that the Eq.~\ref{eq:mode normalization} is no longer satisfied, and for the propagated modes $M_{mm}\le$1 and $M_{mn}\ne$0 for $m\ne n$. Therefore, the coherent modes of the source after propagation and cropping by the beamline are no longer coherent modes of the resulting beam. Fig.~\ref{fig:mode transmission} shows that the most important transfer ratio ($M_{mm}$) transfer for the lowest mode is about $M_{00}$=25\%, then $M_{22}$=10\% and $M_{44}$=6\%. The other modes shown transfer less than 5\% transfer, and modes higher than $m$=10 practically do not contribute to the final spectral density. Note that, as discussed before, the transmission of mode $m$=2 is higher than mode $m$=1, because this one does not display a central lobe (Fig.~\ref{fig:spectraldensity}c).

This analysis of the mode transmission is more dramatic if one looks at the total intensity in the mode, which is given by its eigenvalue. Neglecting modes higher than 10, the lowest coherent mode ($m$=0) carries the 49.5\% of the beam intensity, 18.4\% for the mode $m$=2, and less than 5\% for the others. 


\begin{figure}\label{fig:rediagonalization}
    \centering
        \includegraphics[width=\textwidth]{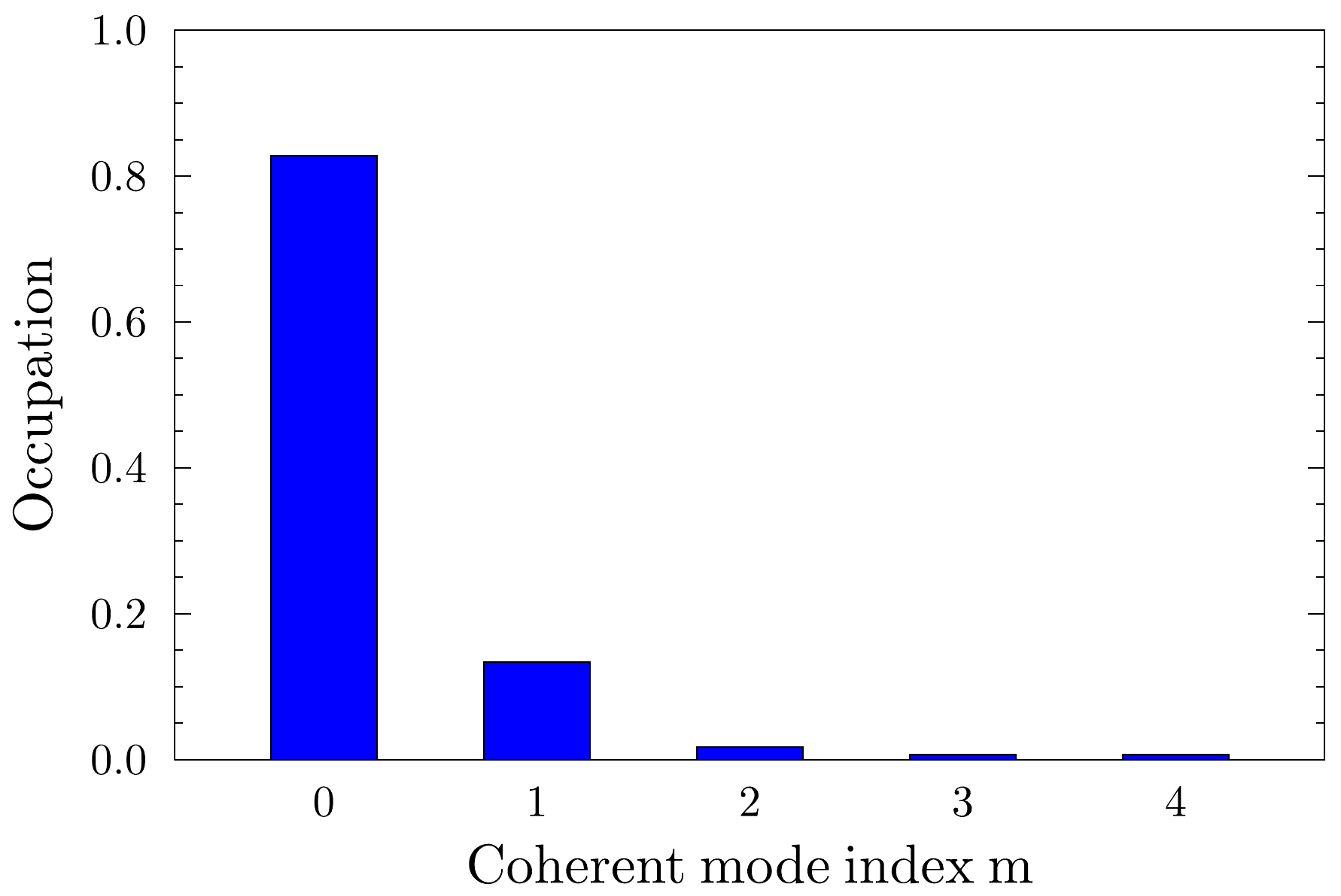}
    \caption{Occupation of the coherent modes of the final focused beam.}
\end{figure}

\begin{figure}\label{fig:final modes}
    \centering
        \includegraphics[width=0.95\textwidth]{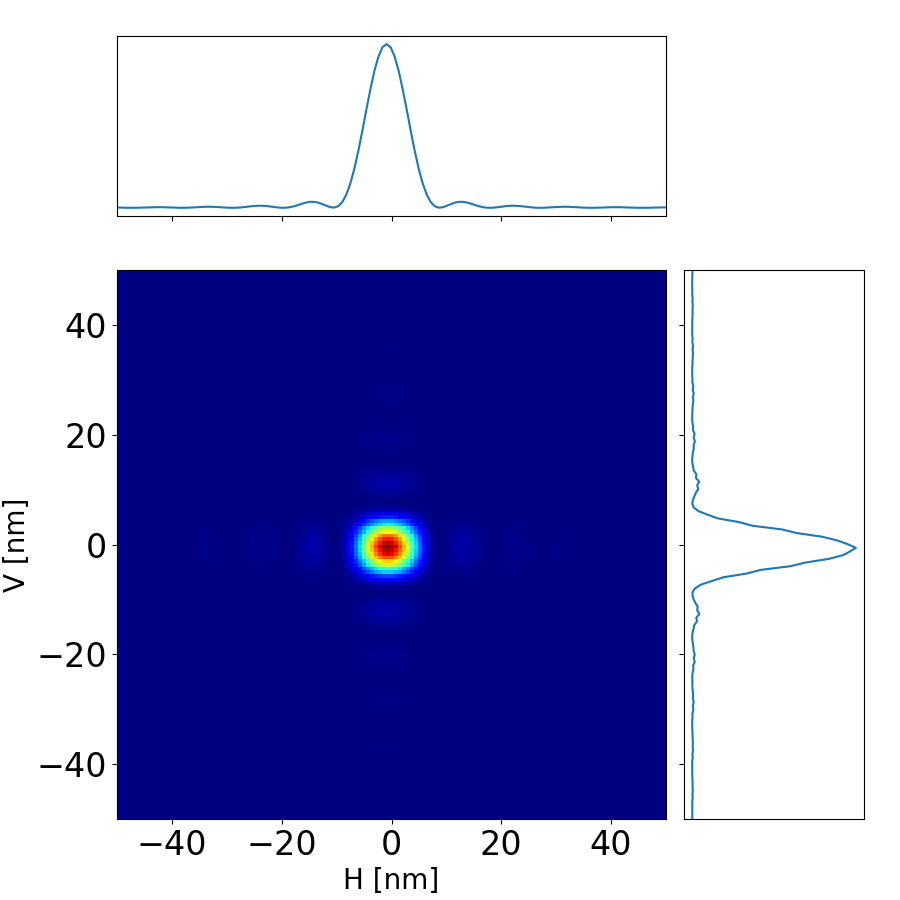}
        \includegraphics[width=0.95\textwidth]{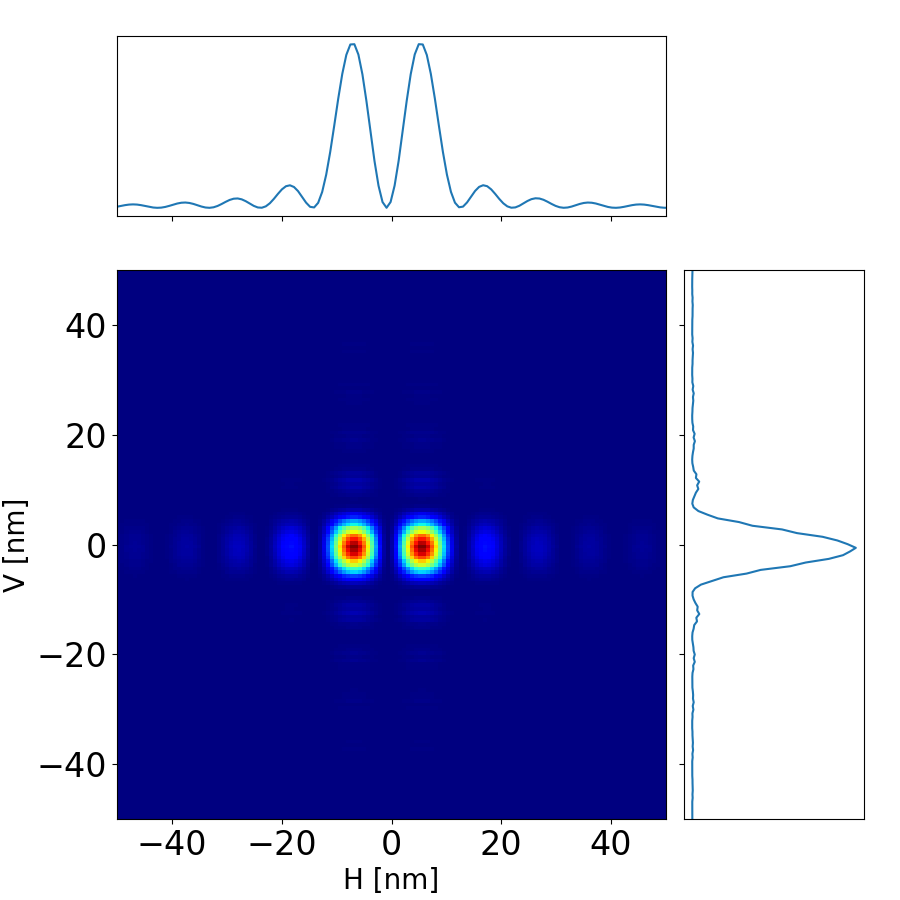}
    \caption{Coherent modes of the focused beam. Top: lower mode ($m$=0), with 83\% occupation and FWHM of 7.3$\times$6.7~nm$^2$. Botton: mode $m$=1, with 13\% occupation and FWHM of 19.0$\times$6.7~nm$^2$}
\end{figure}

As commented before, the coherent modes at the source, propagated and cropped by the beamline, are not longer orthonormal, therefore they cannot be called coherent modes of the resulting beam. At this point, the new coherent modes can be calculated by applying a new coherent mode decomposition on the cross-spectral density of the transmitted beam, which is known from source coherent mode expansion. Making this numerical calculation with COMSYL, one obtains a mode occupation spectrum (see Fig.~\ref{fig:rediagonalization}) where the lowest mode accounts for 83\% of the total spectral density. The next mode accounts for only 13\% and the rest is practically negligible. The intensity of these two coherent modes is shown in Fig.~\ref{fig:final modes}. The first mode has FWHM dimensions of 7.3$\times$6.7~nm$^2$. The spectral density, that is practically given by the contribution of the first two modes has a FWHM equal to the $m$=0 mode because in horizontal the $m$=1 mode does contribute only to the tails and in vertical is identical to mode $m$=0.

This analysis confirms the quantitative predictions discussed in the previous sections, but gives a quantitative figure of the coherence of the beam, a  coherent fraction of CF=83\% that measures the ``coherence purity''.

\section{Discussion} 
\label{discussion}

After the complete analysis done in the previous section on how the ID16A optics modify the characteristics of the X-ray beam, one can verify that the analytical previsions (Section~\ref{level0}) were a good estimator of the values obtained using more sophisticated methods. The size of the focused beam (e.g, 10$\times$5~nm$^2$ for EBS) and the transmission (3\%) calculated analytically are in good agreement with geometric ray-tracing (8.4$\times$4.8~nm$^2$, and 2.28\%). When characterizing beam sizes, one usually assumes that the intensity distributions are Gaussian and therefore they are well characterized by the FWHM value. From Fig.~\ref{fig:ray-tracing} one can see that the horizontal intensity distribution is not Gaussian for the ESRF-1, as a result of cropping the beam with the VSS opened to 50$\mu$m. Although the electron beam statistics is in a very good approximation Gaussian, the undulator emission is not. In our ray-tracing simulations, we always suppose that we work at the resonance, then we applied the Gaussian approximation used in Eqs.~\ref{eq:photon small sigmas} and \ref{eq:photon big sigmas} to compute the geometrical characteristics of the source. This is certainly acceptable at the resonance but will fail out of resonance. Indeed, this lack of Gaussianity is relevant in the calculations of the coherent fraction. We approximate the coherence fraction by Eq.~\ref{eq:coherent fraction}, which for zero emittance ($\sigma_{x,y}=\sigma_{x',y'}$=0) becomes CF=1, as expected because a filament electron beam in a magnetic field emits a fully coherent wavefront. However, the photon source for a filament beam would have an emittance of $\sigma_u \sigma_{u'} \approx \lambda / 2 \pi$ (using Eqs.~\ref{eq:photon small sigmas}), which is in contradiction with theory, that says that a pure coherent state has emittance of $\lambda / 4 \pi$, as happens for a Gaussian photon beam. This is justified considering that Eqs.~\ref{eq:photon small sigmas} are approximations obtained by fitting with Gaussians the theoretical non-Gaussian distributions (see Ref.~\cite{elleaume}). The correct definition of CF as the occupation of the lowest coherent mode does not present these incongruences. However, its calculation requires to perform the coherent mode decomposition, as done in Section~\ref{comsyl} which is very costly. It is, therefore, the case to find the compromise between accuracy and cost, or where in the hierarchical series of methodologies proposed one should stop. 

The combination of ray-tracing and wave optics methods used indicate that the focalized beam size due to  demagnification with mirrors (giving for EBS 10$\times$5~nm$^2$ analytically, 8.8$\times$4.5~nm$^2$ with ray-tracing) has to be corrected with diffraction effect (11.2$\times$8~nm$^2$ hybrid, 12.56$\times$9.38~nm$^2$ SRW) and then add the slope error effect. The slope errors do not blow up the size and kit it in the specifications range (11.4$\times$8.5~nm$^2$ hybrid 13.6$\times$9.14~nm$^2$ SRW). It is important to treat the slope errors correctly. Slope errors can be included in the ray-tracing using the data in Appendix~\ref{appendix:metrology} and applying specular reflection on the optical surface with errors. This is the standard model of including slope errors in SHADOW, using measured profiles or synthetic ones. It works well for the case of incoherent beams, but in our case, because of the high coherence of the beam the ray-tracing method (specular reflection) overestimates the blowing of the focal spot conducting to wrong results. Therefore the slope errors of the KB system have been considered using a ``wave optics approach'', as explained in Ref.~\cite{hybrid}. Here, the mirror height errors $z(x)$ are projected to the plane perpendicular to the propagation ($z(x) \sin \theta$, with $\theta$ the grazing angle). They introduce a phase shift $\psi$ in the wavefront phase proportional to the optical path $\psi = k z(x) \sin \theta $. This is the algorithm used in the simulations in Sections~\ref{level1}, \ref{srw_se} and \ref{srw_me}.


The wavefront simulation of the beamline includes the propagation of the beam from element to element (drift spaces) using different possible propagators. A propagator based on the Freskel-Kirchhoff integral is used. The applications of this (and other) operators imply the calculation of an integral (or sum) at each pixel of the image plane. This is costly from the computer point of view, so Fourier Transforms (Fourier Optics) and its Fast Fourier Transform implementation are used to reducing the number of operations and the calculation time. This is practically the only choice for propagating 2D wavefronts. The use of Fourier Transforms adds another problem: the result is very dependent on the sampling of the wavefront (i.e., pixel size, number of points, the dimension of the window where the wavefront is defined). This means the usually one has to test many configurations before finding one that works reasonably well (even though one can never guarantee the complete accuracy of the sampling used). A full wavefront optics simulation is an iterative process, where the user has to refine the sampling parameters element by element to be sure the wavefront at each position is sampled correctly. This is a time-consuming work in particular for 2D simulation. We found very useful to start a wavefront simulation in 1D, separating the horizontal and vertical simulations, using a simplified system with ideal sources and ideal focusing elements. The use of 1D propagation is justified because in synchrotron beamlines the horizontal and vertical coupling is usually small. These 1D simulations help in refining the choice of propagator and its parameters (resampling, zoom factors) that can be reused in full 2D simulations.

Regarding computational effort, simulations concerning ray-tracing (Section~\ref{level1}) and coherent optics (Sections~\ref{wofry} and \ref{srw_se}) can be done interactively in a normal laptop. However, for those concerning partial coherence, very long calculations are needed. For example, Monte Carlo calculations in Section~\ref{srw_me} for the multi-electron case spent a few hours on a 56-core computer cluster for simulating 22000 electrons. For performing the COMSYL coherent mode decomposition the calculation time depends very much on the dimensions of the problem. For EBS the wavefronts were sampled using 1007$\times$335 pixels, and 1103 modes were computed representing 98\% of the spectral density. This represents 5 MB per wavefront and a matrix of 1695 GB to diagonalize. It took less than four hours in a cluster of 28 cores. The calculation is more expensive if one wants to decompose a source that is pretty incoherent like our ESRF-High-$\beta$. This is a limit case, it was used wavefields of 10075$\times$201 pixels to accommodate the large spatial extension of the source. We obtained 4016 modes totalizing only 86\% of the spectral density.  The calculation took more than 4 days in an 84 core cluster. This was done as a test case, to demonstrate the feasibility of the calculations. It is not recommended to do similar calculations that imply a cost because a source that is quite incoherent was treated with partial coherence methods. 

Some effects that are typically studied when designing and simulating a beamline have not been discussed here. For example, the study of the power emitted by the source and its heat load effect in the first (or firsts) slits and optical elements (here the ML monochromator). Also, the reflectivity of the elements is not considered. The simulation of this was not included in the simulations, because it can be estimated analytically considering an average reflectivity about 0.7 for multilayers and 0.9 for mirrors, therefore a transmission factor of 0.44 for the beamline flux due to the reflectivity of mirror coatings. 


\section{Summary and conclusions}
\label{summary}

We have obtained quantitative values of flux, transmission, size of the focused beam, and coherent fraction for ID16A, a modern beamline with already exceptional performances in terms of nanofocusing and coherence. We predicted advances in the improvements in these parameters when the EBS will start in 2020. We obtained these parameters using a hierarchical flowchart applying methods with increasing complexity and requiring a higher computational effort. 

The analytical estimation of the main parameters is the mandatory first step in the beamline conception and analysis. We applied concepts based on optical magnification and numerical aperture acceptance and calculated approximately the coherent fraction at the source. These values help to perform a sanity check of the computer simulations. A ray-tracing simulation is inexpensive and provides important information. The comparison of the CF with the beamline transmitivity would give an idea of the beam coherence, reflected in the number modes present in the final beam. For high coherence, like for ID16A, the wave optics calculations are necessary. If the final beam has a very high coherence coherent optics (1D simplifications, and single-electron calculations) may give good results. Full calculations based on coherent mode decomposition Were introduced here. It is a new way of quantitatively assess the quality of the partially coherent beam via the calculation of the coherent modes of the source, propagation of the modes and recalculation of the coherent mode expansion on the final beam. The CF or occupation of the first mode is the parameter used to measure the coherence quality.

%


\appendix

\section{Metrology of the KB mirrors}
\label{appendix:metrology}
\begin{figure}\label{fig:metrology}
\includegraphics[width=0.95\textwidth]{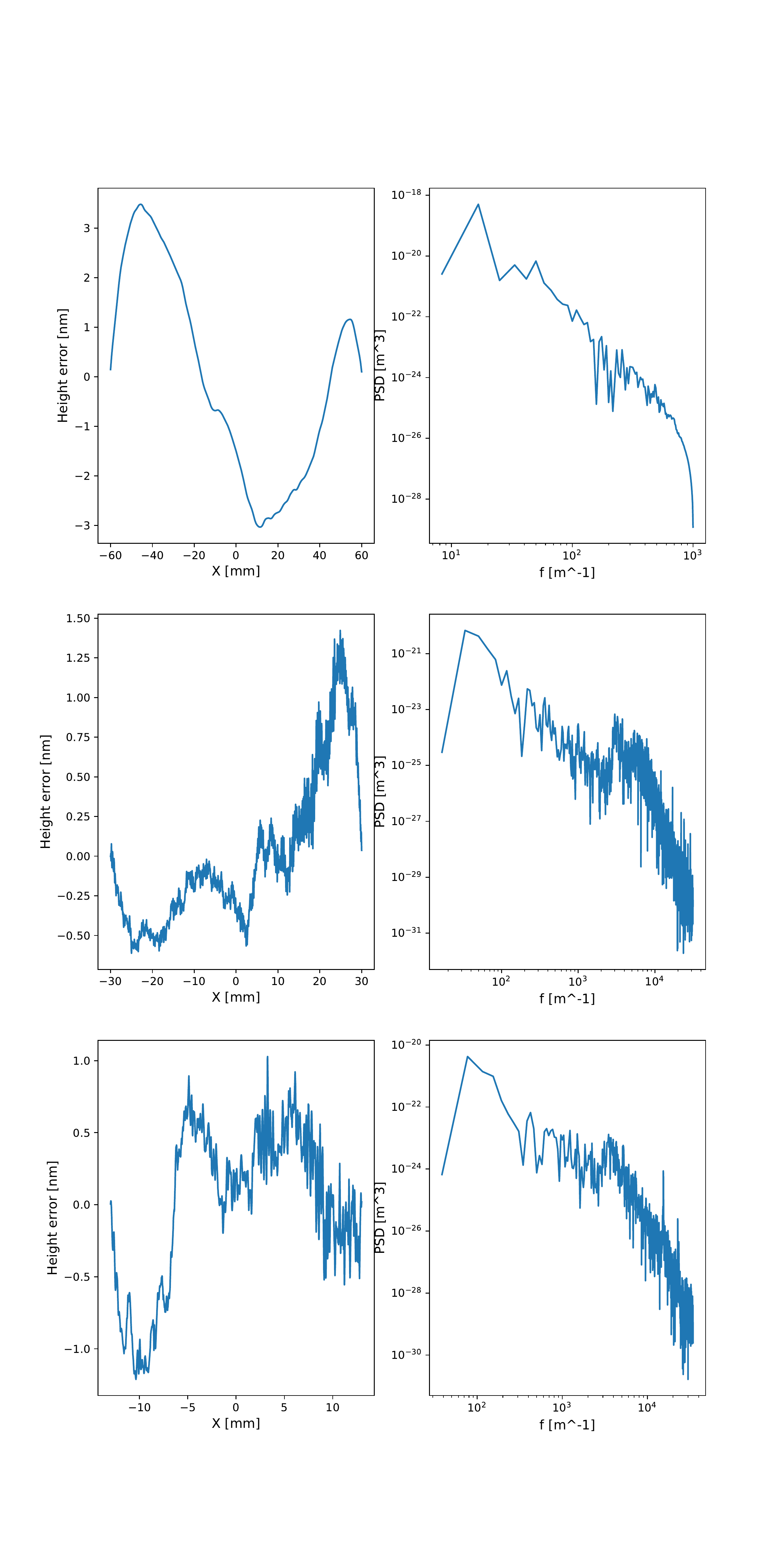}
\caption{Error height profiles used for the simulations (left) and power spectral density (right) for the ML monochromator (top row), the KB vertical focusing mirror (central row), and the KB horizontal focusing mirror (bottom row).}
\end{figure}

The residual height error of the multilayer monochromator and the KB mirrors, after removing the main ellitical shape, together with their power spectral densities are shown in Fig.~\ref{fig:metrology}. These profiles are used along the text in ray-tracing and wave optics simulations.

\bibliography{paper-hierarchical}


\end{document}